\documentclass[aps,prl,groupedaddress,notitlepage]{revtex4-1}
\usepackage{fullpage}
\usepackage{graphicx} 
\usepackage{amsmath}
\usepackage{amsfonts}
\usepackage{amssymb}
\usepackage{hyperref}
\usepackage{float}
\usepackage{mdframed}
\usepackage{color}
\usepackage{comment}

\definecolor{explanationcolor}{RGB}{240,240,255}
\definecolor{examplecolor}{RGB}{240,240,240}
\definecolor{exercisecolor}{RGB}{250,240,240}
\definecolor{problemcolor}{RGB}{225,255,240}
\definecolor{challengecolor}{RGB}{255,240,225}
\definecolor{relevlitcolor}{RGB}{250,250,250}
\definecolor{hyperlinkcolor}{RGB}{0,0,0}
\definecolor{hypercitecolor}{RGB}{0,180,90}
\usepackage{hyperref}

\newenvironment{explanationbox}{\begin{mdframed}[backgroundcolor=explanationcolor,linewidth=1pt]}{\end{mdframed}}
\newenvironment{examplebox}{\begin{mdframed}[backgroundcolor=examplecolor,linewidth=0pt]}{\end{mdframed}}
\newenvironment{exercisebox}{\begin{mdframed}[backgroundcolor=exercisecolor,linewidth=0pt]}{\end{mdframed}}

\newcounter{explanation}
\def\theexplanation{\arabic{explanation}}

\newcounter{example}
\def\theexample{\arabic{example}}

\newcounter{exercise}
\def\theexercise{\arabic{exercise}}

\newcounter{problem}
\def\theproblem{\arabic{problem}}

\begin{document}


\author{Anupam Sengupta}
\email{anupam.sengupta@uni.lu} 
\affiliation{Physics of Living Matter, Department of Physics and Materials Science, University of Luxembourg, 162 A, Avenue de la Faïencerie, L-1511 Luxembourg City, Luxembourg}


\title{Planktonic Active Matter}

\maketitle

\section*{Introduction}

Plankton, a fundamental component of our biosphere, comprises organisms spanning a wide range of dimensions, morphology, functional, and behavioural traits. Functionally, they can be broadly categorized into phytoplankton, zooplankton, mycoplankton, bacterioplankton, and virioplankton, depending on their position in the food web and ecological functions \cite{kiorboe2018book}. Phytoplankton--the focus of this chapter--is one of the most important functional groups made up of light-harvesting prokaryotic or eukaryotic organisms. They photosynthetic microorganisms which are at base of nearly all aquatic food webs. As a key player of the ocean and freshwater ecosystems, phytoplankton impact global biogeochemical cycles, produce close to half of the world’s oxygen, and are important sources of algal biofuel. The word, phytoplankton, is a portmanteau of the Greek words, \textit{phyton} meaning plant, and \textit{planktos}, signifying a wanderer or passive drifter. For long phytoplankton have been considered to be passive drifters, their spatio-temporal locations determined largely by the environmental fluid flows. However, decades of satellite-, field- and lab-based studies have confirmed that the movement of phytoplankton can occurs actively, and they may not be wandering - horizontally or vertically along the water column - due to the fluid flows alone \cite{kessler1986, siegel1998, elgeti2015}.

\begin{figure}[h!]
	\begin{center}
		\includegraphics[height=8cm]{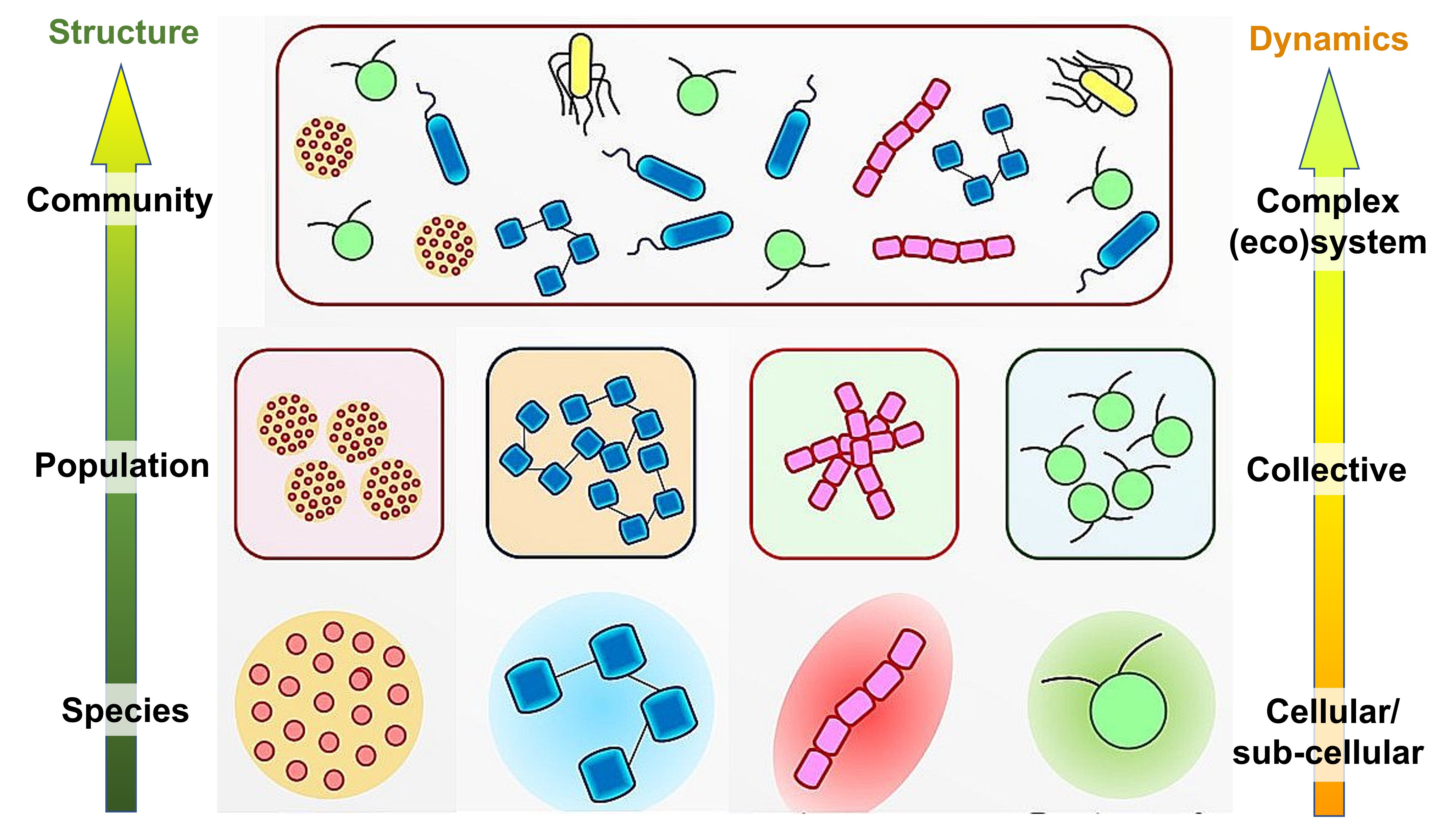}		
	        \caption{
	        {\bf Scales and complexity of planktonic active matter.} Planktonic active matter presents an emergent system across different scales: individual, population and community; and complex dynamics at sub-cellular and cellular to collective and ecosystem scales. The cross-scale active matter system responds to both abiotic (temperature, fluid flow and light conditions) and biotic factors (nutrients, pH, secondary metabolites) characteristic of the relevant ecosystems. Active modulation of cell phenotypes, including morphology and motility, enable planktonic microorganisms to interact with other individuals and species, and rapidly adapt to changes in their micro-environment. This multi-scale approach allows us to apply an active matter framework for understanding natural ecological systems and their emerging structure and functions due to changing conditions.  }\label{fig:scales}
	\end{center}
\end{figure}

Planktonic active matter represents a highly diverse community of organisms, with hierarchical complexity in their composition, structure, and dynamics ranging from intraspecific (within a given species) to inter-species and species-environment interactions (Figure \ref{fig:scales}). Planktonic active matter spans orders of magnitude in size (sub-micron to sub-millimeter range), distinct morphologies, ecosytem-dependent photosynthetic activities, and biological functions. Light, together with nutrients and turbulence determine the ecology of phytoplankton \cite{margalef1978mandala, sunagawa2015,devargas2015}, thus making their position along the vertical water column highly consequential. Phytoplankton occupy the so-called \textit{photic} zone - the light-rich region of the aquatic ecosystems. Light is a key determinant phytoplankton physiology and fitness, and mediates vital functions by regulating endogenous circadian cycles of light-harvesting bacteria and algae \cite{blankenship2021book}. Many species of phytoplankton are motile , $i.e.$, they can actively migrate through the water column by leveraging \textit{gravitaxis}, the directional movement in response to gravity \cite{Braun2018gravisensing}. Gravitactic phytoplankton cells migrate upward - against the direction of the gravity force and toward light - during the day, and change their swimming direction downward - toward higher inorganic nutrient concentrations - at night \cite{wada1985, sullivan2003turbulence,kotar2020dvm,tergolina2021}. While the size of individuals undertaking this daily vertical migration (DVM) is miniscule, the sheer number of cells involved makes DVMs some of the largest and most important concerted microbial migrations on Earth. DVMs contribute significantly to the sequestration of carbon from the atmosphere to the deep ocean, among others through sinking particulate organic matter. Groups of motile gravitactic species, including dinoflagellates and raphidophytes, are frequently found to generate harmful plankton blooms, or more commonly, harmful algal blooms (HABs). A complex interplay of cell motility and morphology, alongside abiotic and biotic factors like the ambient temperature, fluid flow, nutrient concentrations, pH, and seasonal factors drive the formation of the HABs \cite{mcgillicuddy2007, behrenfeld2014}. 

Complementing motile species, a second important group of planktonic microorganisms comprises the diatoms \cite{pierella2020}. Unlike their motile counterparts, diatoms generally lack appendages (for e.g., the cilia or flagella) required to generate propulsion forces. Their ecology, as a consequence, is tightly coupled to the local fluid flows, which act as conveyors for moving cells from one point to another. Conversely, motile species dominate the calmer regions of the ocean, while diatoms are frequently associated with highly dynamic settings, for instance in the regions of high turbulence. Along the vertical water column, diatoms leverage an array of biophysical mechanisms to either maintain or alter their buoyancy, allowing them optimal access light and nutrients. Diatoms have evolved different modes to regulate density, including active replacement of heavy ions within vacuole (triggred by light, nutrients or osmotic stress \cite{falciatore2000}, reduction of the starch or carbohydrate inclusions \cite{richardson1995,moore1996}, or over longer timescales, by 
bio-silicification \cite{raven2004}. Remarkably, diatoms show a rapid control of buoyancy under nutrient-limited conditions, thereby potentially increasing the diffusive transport of nutrient molecules to the cell-surface \cite{gemmell2016}. Physiologically, diatoms have been found to access nitrogenous compounds more efficiently (at lower concentrations), thereby exhibiting relatively higher photosynthetic and growth rates compared to the motile dinoflagellates \cite{hinder2012}.

Planktonic active matter demonstrates exquisite mechanisms to diversify their biophysical traits in response to a range of physico-chemical cues. The ability of phytoplankton to adapt their motility traits - over both short (within a division time-scale) and longer time scales (spanning multiple generations) - is underpinned by morpological pliability, intracellular reconfigurability, or modulation of the flagellar beating. Diversification of traits, both physiological and behavioral, enables phytoplankton populations to respond, adjust and adapt to changes in their environmental conditions, thus maximizing chances of survival \cite {sengupta2017,carrara2021,sengupta2020,sengupta2022}. This offers a highly rich test-bed for biophysicists to test hypotheses, and drive the field of active and living matter forward. More importantly, understanding how phytoplankton adapt and develop strategies to survive the rapidly evolving nutrient, turbulence and light conditions of today’s oceans remains a crucial challenge. Accounting for the active mechanisms and emergent properties, observed frequently across all planktonic systems, could allow accurate predictions of planktonic community compositions, structures and dynamics across scales and complexities, ultimately advancing the existing models of biogeochemical cycles and biological pumps for today’s aquatic ecosystems.

\section{Physical ecology of plankton}\label{ecology}

\begin{figure}
	\begin{center}
		\includegraphics[height=6cm]{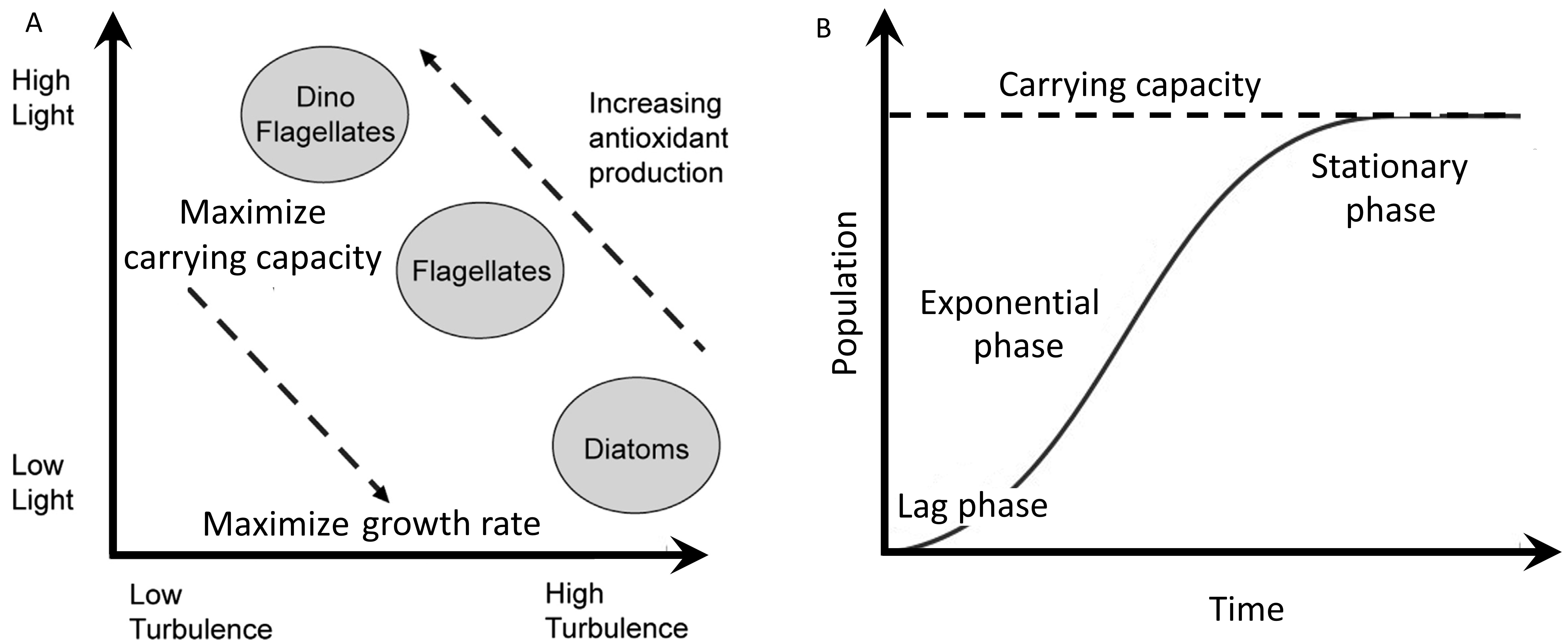}		
	        \caption{
	        {\bf Physical ecology of planktonic microorganisms.}
	   	{\bf a} Depiction of the Margalef's mandala showing the relationship between environmental factors (light and turbulence) and phytoplankton type and antioxidant production in natural habitats. Flagellated species typically occupy low turbulence and high light regions, whereas non-flagellated species, like diatoms, are typically found in regions of high turbulence and low light regions. Adapted from Ref. \cite{allen2011mandala}. {\bf b} The  logistic curve captures growth phases of planktonic species growing under laboratory environments: lag phase, exponential phase and the stationary phase. The specific growth rate of a species is calculated during the exponential phase. The carrying capacity quantifies the average steady state population size under given environmental conditions (representative of a particular habitat), including nutrient availability, light and turbulence conditions, and prey-predator interactions.
	    }\label{fig:mandala}
	\end{center}
\end{figure}

Ramon Margalef, one of the founding fathers of modern marine biology, was among the first to propose and formalize the dependence phytoplankton physiology in relation to their environmental settings, specifically, the levels of light, nutrient concentration and turbulence \cite{margalef1978mandala, allen2011mandala, kemp2018mandala}. In the context of gravitaxis, ambient fluid flows have a direct impact on phytoplankton behaviour and physiology. While strong turbulence can be detrimental to motile phytoplankton, potentially leading to the flagellar or body wall damages, triggering enhanced physiological stress, and reduced growth \cite{carrara2021}. Planktonic species leverage evolutionary coping mechanisms to tackle environmental stressors, by devising adaptive strategies based on the intrinsic plasticity of their functional traits \cite{sullivan2003turbulence, smayda2010, sengupta2017}. These include minute but rapid adjustments of the cell morphology to adjust the swimming stability \cite{sengupta2017}, to reduction of spine length to promote sinking by the dinoflagellate \textit {Ceratocorys horrida} \cite{zirbel2000}, or the formation of chains by bloom-forming \textit{Alexandrium catenella} to adjust their swimming behavior in response to hydrodynamic shear \cite{karpboss2000}. However, fluid flows can be beneficial for non-motile diatom species, specifically due to the enhancement of the encounter rates between cells and nutrient molecules \cite{kiorboe2018book, jumars2009}. Interestingly, alteration of cell morphology may further enhance the access to nutrients, for instance when single cells transform into chain-like morphology observed in many species \cite{musielak2009}. In addition to the cell length, increase of the chain rigidity enhances relative nutrient fluxes, suggesting a critical advantage conferred by the silica frustules often found in diatoms \cite{young2012}.

The different phytoplankton life-forms captured in Margalef's \textit{mandala} (Figure \ref{fig:mandala}\textbf{a})  are based on their ability to adapt and survive in unstable and turbulent environments. Their small size combined with rapid turnover times make external energy input from turbulence a key determinant of phytoplankton fitness. In it's simplest form, fitness can be measured in terms of the growth rates and the carrying capacities of the species (Figure \ref{fig:mandala}\textbf{b}), though, technically, fitness could include multiple associated metrics, including a species' ability of risk- or predator-avoidance, stress amelioration, or maintenance of basic metabolic processes under limited resources. Secondary factors like grazing by predators may be further introduced into the \textit{mandala}, overlaying them with the light (closer to the air-water interface) and nutrients (at depths in the water column) as the other two primary axes which govern phytoplankton eco-physiology.

\begin{figure}
	\begin{center}
		\includegraphics[height=8cm]{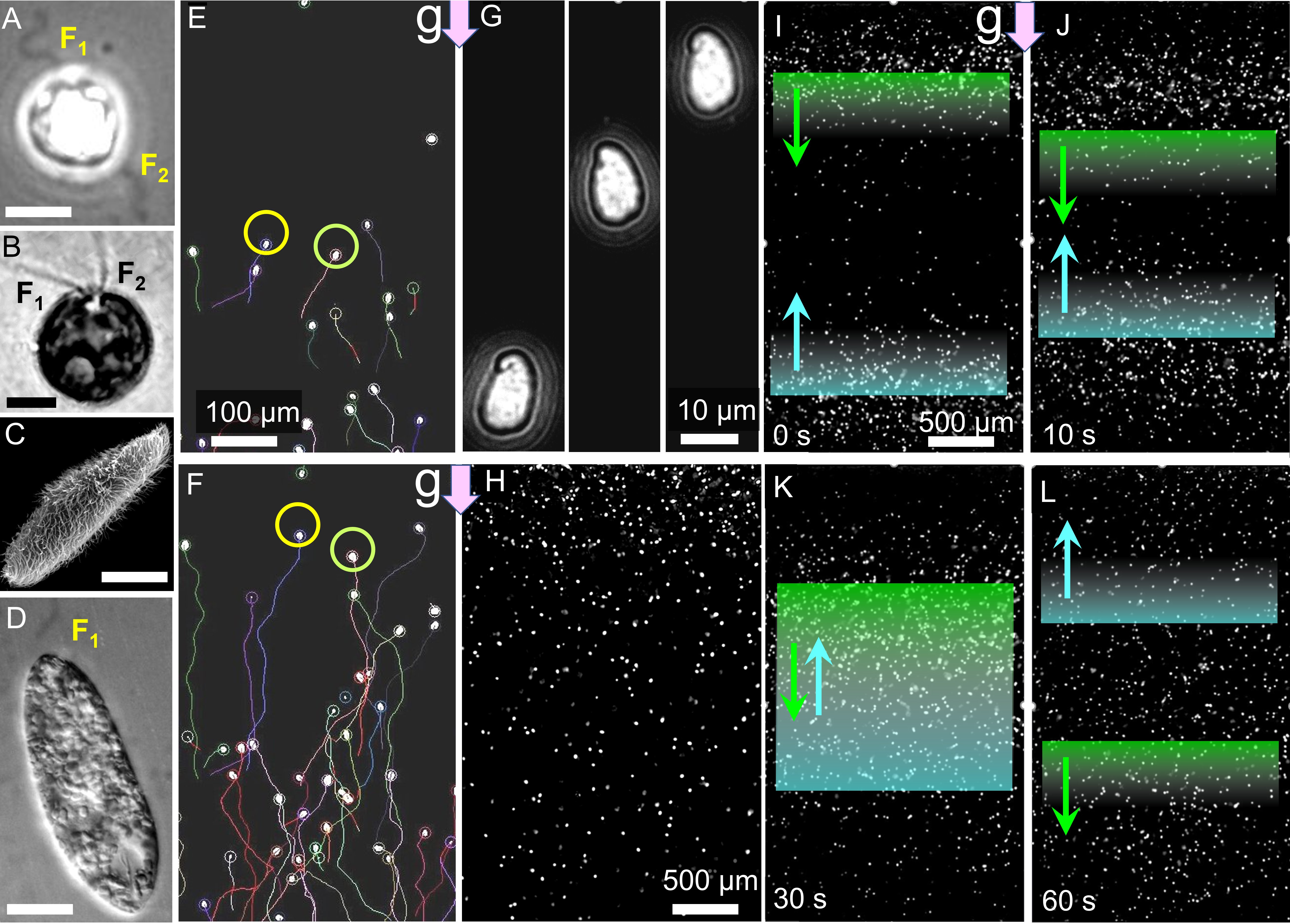}
	\caption{
	{\bf Gravitaxis and diel vertical migration.} 
	Gravitactic species {\bf a} \textit{H. akashiwo}, {\bf b} \textit{Chamydomonas reinhardtii} (image adapted from Ref. \cite{schroda}), both imaged in phase contrast mode, {\bf c} scanning electron micrograph of ciliate \textit{Paramecium tetraurelia} \cite{valentine2012}, and {\bf d} brightfield image of \textit{Euglena gracilis} \cite{hader2022euglena}. $F_{1}$ and $F_{2}$ indicate the flagella present in the corresponding species. {\bf e} and {\bf f} show a poluation of negatively gravitactic species swimming against the gravity direction (downward arrow). The trajectories of the swimming cells are presented using different hues. {\bf g} Single microplankton cell executing gravitactic motion in a vertical column. Cells typically rotate about their body axis while executing a helical swimming trajectory. {\bf h} Stationary distribution of gravitactic phytoplankton in a vertical chamber. A higher cell concentration is observed around the top of the chamber, than at the bottom region of the chamber. {\bf i-l} A sequence of micrographs captures the motion of co-existing negative (up-swimming cells) and positive (down-swimming cells) gravitactic sub-populations in a vertical chamber.  
	}\label{fig:dvm}
	\end{center}
\end{figure}

Phytoplankton frequently encounter diverse fluid dynamic environments: under natural environments which they inhabit, within engineered confinements such as algal bioreactors, or as self-organized emergent flows as in a bioconvecting plume \cite{guasto2012,bees2014maths,bees2020}. Large scale turbulent structures typically cascades to eddy structures at smaller scaler, until the turbulent energy is dissipated by viscosity alone \cite{kiorboe2018book}. Under typical marine conditions, the mean dissipation rate of turbulent kinetic energy varies between $10^{-9} < \epsilon < 10^{-5} ~W kg^{-1}$, corresponding to a Kolmogorov scale, $\eta_{k} = (\nu^{3}/\epsilon)^{1/4}$ = 0.5 – 5 mm; $\nu \approx 10^{-6} ~m^{2}s^{-1}$ being the kinematic viscosity of sea water \cite{estrada1997,jumars2009,sutherland2013}. The Kolmogorov scale represents the smallest scale below which the eddy flow breaks down into a simple linear variation in the fluid velocity. The corresponding order of magnitude of the velocity gradient, given by the Kolmogorov shear rate, $S_{K} = (\eta/\nu)^{1/2}$, thus falls between 0.03 - 3 $s^{-1}$, with its inverse $\tau_{K}= 1/S_{K}$ representing the characteristic time scale of the velocity fluctuations. With a vast proportion of the phytoplankton species smaller than the characteristic Kolmogorov scale of oceans, cells perceive turbulence as an instantaneous, linearly varying fluid velocity across the cell body \cite{barry2015}. The interplay between phytoplankton motility and ocean turbulence has drawn considerable attention, leading to a deep understanding of both the biophysical and biomechanical aspects, particularly of the vertical migration. Numerous phenomena have been reported including the formation of thin planktonic layers due to shear flows \cite{durham2009}, emergence of phytoplankton clusters, also referred to as \textit{patchiness}\cite{durham2013,lillo2014,breier2018patch}, and the enhancement of the vertical migration of chain-forming phytoplankton through turbulence \cite{lovecchio2019}. More recently, it has been shown that major groups of motile phytoplankton (raphidophytes and dinoflagellates) harness active, behavioral changes in response to turbulent cues \cite{sengupta2017}. The response - manifested as morphological transformations which impact the stability of the swimming cells - is underpinned by the generation of the stress markers, called the Reactive Oxygen Species (ROS) \cite{carrara2021}. Furthermore, morphological transformations can occur under nutrient-limited settings, which together with the growth and intracellular translocation of energy-storing lipid droplets, can govern migratory strategies of phytoplankton in a species-specific manner \cite{sengupta2022}.


A vast majority of the planktonic microbes perceive gravity, and use it to adjust their position in the water column, in combination with other external cues like light and chemical gradients (Figure \ref{fig:dvm}). Directed movement along or against the gravity direction is called \textit{gravitaxis} and can be positive (downward swimming) or negative (upward swimming) \cite{roberts2006,Braun2018gravisensing}. The execution of the nagative or positive gravitaxis - observed across raphidophytes, dinoflagellates, and ciliates (Figure \ref{fig:dvm} \textbf{a-d}) depends on a range of factors, the primary among them is the circadian rhythm \cite{lakin-Thomas2004, schuech2014}. In addition, some ciliates and flagellates are able to perform \textit{gravikinesis} whereby cells modify their swimming behaviour by activating a kinetic response: they speed up during the upward swimming phase, and decelerate during downward swimming \cite{machemer1992, hemmersbach2002}. This kind of motion, observed typically in larger organisms, depend on the local environmental conditions and enable cells to compensate sedimentation rates, either partially (for instance, in \textit{Paramecium} \cite{machemer1991}) or fully (like in \textit{Tetrahymena} \cite{kowalewski1998}). Planktonic microbes have an array of different mechanisms to perceive gravity, including the \textit{statolith}: a heavy bio-mineralized organelle that presses onto the cell's gravireceptor; mechanosensitive ion channels: they act as gravireceptors which amplify changes in the gravity (or acceleration) based on which the direction of swimming is altered; and physiological stress markers like the ROS \cite{Braun2018gravisensing, sengupta2020}.


\subsection{Morphology and organelles of phytoplankton}

\begin{figure}[b!]
	\begin{center}
		\includegraphics[height=5cm]{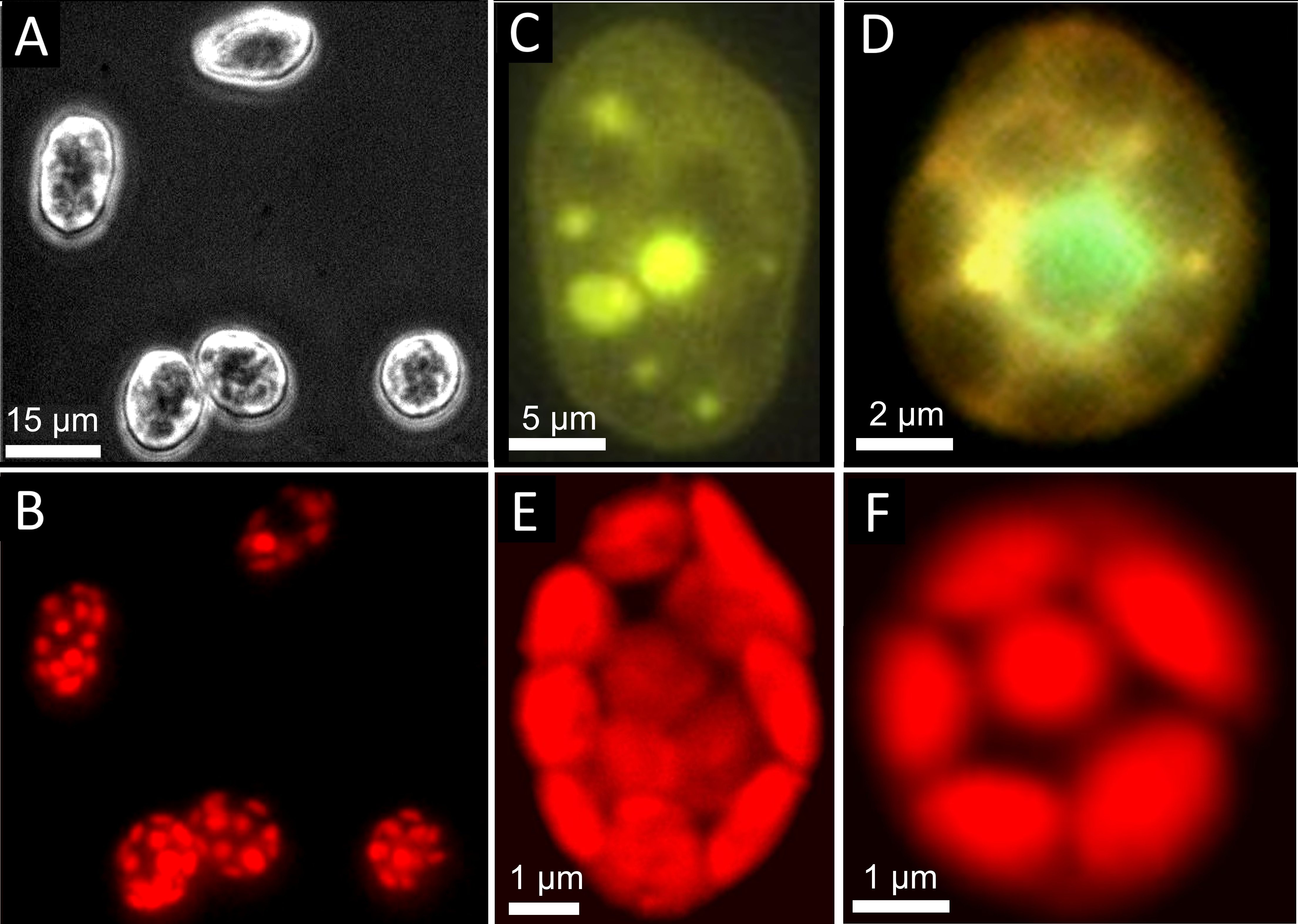}		
	        \caption{
	        {\bf Cell morphology and functional  organelles.} 
	   {\bf a} Phase contrast image of raphidophyte \textit{Heterosigma akashiwo}, a motile biflagellate microplankton species. {\bf b} Corresponding autofluorescent micrograph of \textit{H. akashiwo}, illuminated by blue light, reveals the light-harvesting organelles called the chloroplasts. {\bf c} Epifluorescence microscopy of intracellular lipid droplets (yellow-green hue) stained using the dye Nile Red. {\bf d} Dual-channel epifluorescent imaging capture the cell nucleus (green hue, stained using Syto9 stain) and lipid droplets (yellow-green hue, stained using Nile Red dye). {\bf e} and {\bf f} Magnified view of chloroplasts embedded on the outer cell membrane of \textit{H. akashiwo} cells, visualized using autofluorescence imaging. 
	    }\label{fig:MD:organelles}
	\end{center}
\end{figure}

The morphology and size of phytoplankton have been found to play a crucial role on the growth, uptake and survival of species. Across different scales and organizational complexity, body size correlates with various traits of species, impacting the composition, structure and dynamics of the phytoplankton networks and food webs, and their stability and resilience to perturbations \cite{woodward2005}. In regards to the growth rates, larger motile cells have been reported to outperform motile cells of smaller diameters, possibly due to the relatively higher nutrient uptake rates in larger cells, particularly under turbulent conditions \cite{kiorboe2018book, guasto2012, fraisse2015}. Under nutrient-limited settings, larger species are often found to outperform the smaller ones \cite{cozar2005}, while cell morphology governs the the active hydrodynamic strategies which the cells put in use to navigate their fluidic settings \cite{padisak2003, sengupta2022}. As turbulent strength increases, shear forces can be detrimental for the motile species, with even complete cessation of motility \cite{sengupta2017, carrara2021}. Furthermore, in regards to the nutrient uptake, a threshold is attained, size no longer offers a competitive advantage to the swimming cells. For pico- and nanoplankton, the impact of cell morphology on the their eco-physiology has been reported to be insignificant. Owing to the dependence of the metabolic constraints with body-size scaling, investigations are currently underway to develop a quantitative framework within which such microscale traits can be incorporated toward description of the structure and functioning of populations and networks at larger scales. The quantitative roles of cell morphology under turbulent cues and nutrient limitation is taken up later in the chapter. In addition to cell size and morphology, a growing body of recent literature has demonstrated the role of intracellular organelles in governing the swimming properties (Figure \ref{fig:MD:organelles}). These include the cell nucleus (the heaviest organelle in a cell), energy-storing lipid bodies, carbohydrate and starch reserves, chloroplasts and gas vacoules \cite{sengupta2017, sengupta2022, milo2016book}. By altering the physical size, density, and position with the cell body, cellular organelles can alter the swimming speed, orientational stability, velocity correlations (switching from ballistic to diffusive swimming or vice-versa), thereby impacting the overall motility characteristics of the swimmers. 

   \begin{figure}
	\begin{center}
		\includegraphics[height=5cm]{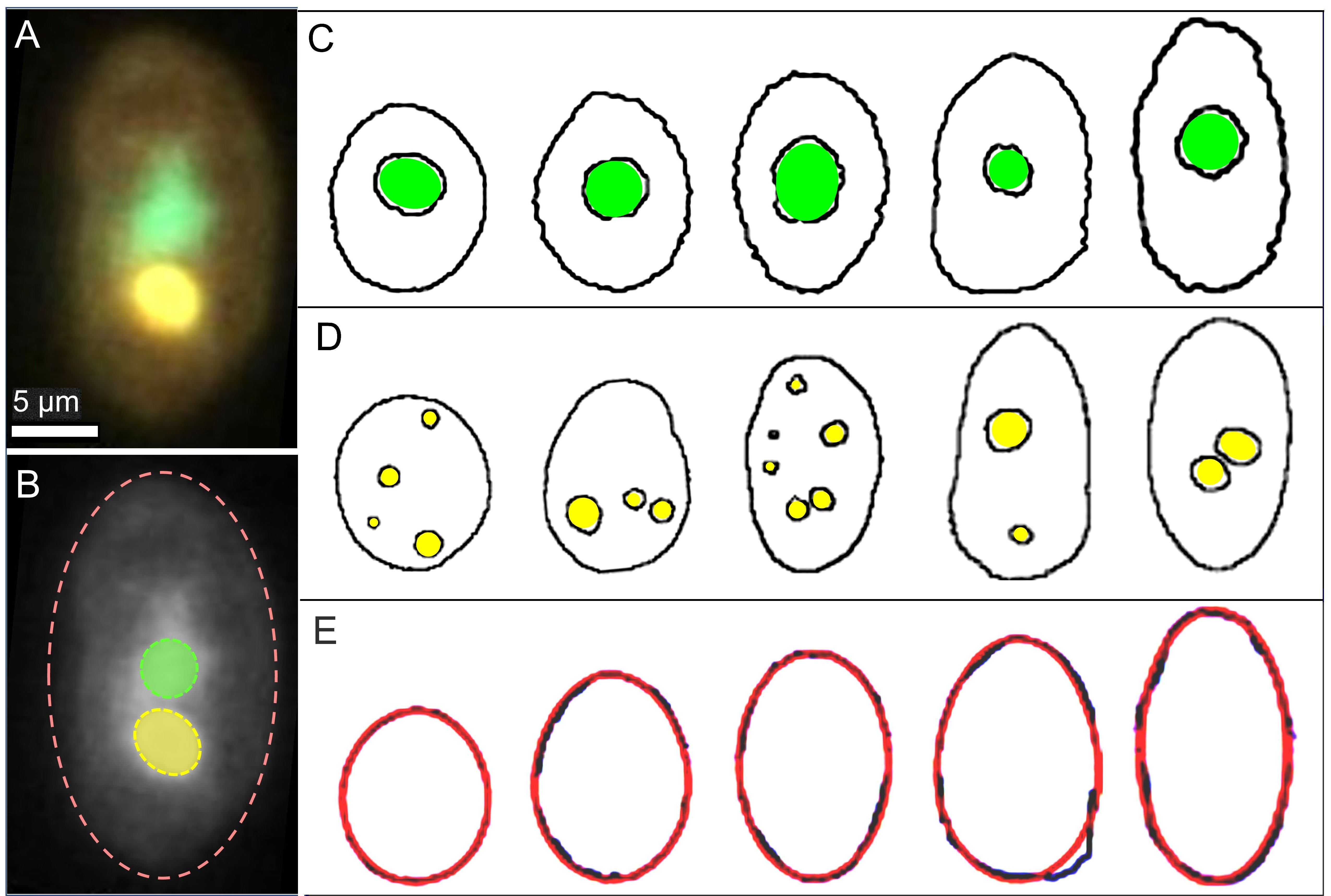}		
	        \caption{
	        {\bf Quantifying cell morphology and intracellular organelles.} Raw {\bf a} and grey-scale {\bf b} images showing the cell morphology and intracellular organelles within a microplankton. Dashed lines denote the contours of the cell body and the organelles. The nucleus (green) and lipid droplet (yellow) were simultaneously visualized using epifluorescence imaging. Image analysis can be used to extract the contour of cell and the positions of the {\bf c} nucleus and {\bf d} lipid droplets. The experimentally extracted cell contours are fitted with a three-parameter curve. Adapted from Refs. \cite{sengupta2017, sengupta2022}.
	    }\label{fig:morphquant}
	\end{center}
\end{figure}

Using bright-field, phase-contrast, and fluorescent-based single cell imaging techniques, precise detection and quantification of the cell morphology and organelles can be made. As shown in Figure \ref{fig:morphquant}, raw experimental images can be analyzed by image processing tools to extract the cell and organelle contours, sizes, and ultimately fitted to mathematical functions to obtain various feature dimensions \cite{sengupta2017,sengupta2020,sengupta2022}.

\subsection{Encounter rates and kernels}

Planktonic active matter frequently encounter different molecules which make up their micro-environment, and with one another, either stochastically or while executing prey-predator dynamics. The \textit{encounter rate kernel} \cite{visser2006}, a measure of clearance rate (or zone of influence) of a particle or a cell, depends on a number of factors: the mobility of the cell (static, swimming or sinking), dynamics of the surrounding fluid (stationary versus flowing conditions), laminar versus turbulent eddies, and the shape (and symmetry) of the cell \cite{kiorboe2018book,andersen2019, arguedas2022}. Encounters between microoganisms play a crucial ecological role, from predator–prey interactions, to food-web structures and optimal foraging \cite{sundby1997,titelman2003,dolger2017}, and finding mates in larger plankton like the copepods \cite{gerritsen1977, dusenbery2009book}. Following Ref. \cite{kiorboe2018book}, the encounter rate between cells in a given population can be written as:

\begin{equation}\label{eqn:encounter}
\begin{gathered}
E = \beta C_{i}C_{j} = \beta C_{1}^{2}
\end{gathered}
\end{equation}

where $i=j$ holds for same species, and $i \neq j$ for different species (e.g., in a prey-predator interaction); $\beta$ being the encounter rate kernel. The concentration of cells over time can be written as:   

\begin{equation}\label{eqn:encounter_growth}
\begin{gathered}
\frac{dC_{1}}{dt} = -\beta C_{1}^{2}
\end{gathered}
\end{equation}

By solving the concentration equation, one can obtain the growth dynamics of cells aggregating, or extend this further for other relevant settings (by appropriately changing $i$ and $j$ values). The corresponding kernels (also known as coagulation kernels, with dimension of volume rate) for different mechanisms (thermal diffusion, settling particles with different speeds, and turbulence driven encounters) are given as:

\begin{gather}
4 \pi(D_{i}+D_{j}) : \text{Brownian diffusion} \\
0.5 \pi a_{i}^{2} \lvert{u_{i}-u_{j}\rvert} : \text{Differential settling for $a_{i} \leq a_{j}$} \\
1.3 \gamma(a_{i}+a{j})^{3}E_{i,j} : \text{Small-scale turbulent shear} 
\end{gather}

where, $D_{i}$, $a_{i}$ and $u_{i}$ represent the diffusion coefficient, dimension and speed respectively; $\gamma$ is the sub-Kolmogorov shear rate; and  

\begin{equation}\label{eqn:encounter_turb}
\begin{gathered}
E_{ij} = 1 ~\text{for $a_{i} = a_{j}$} \\
E_{ij} = \frac{7.5(a_{i}/a_{j})^{2}}{\left[1+2(a_{i}/a_{j})\right]^{2}} ~\text{for $a_{i} < a_{j}$}
\end{gathered}
\end{equation}

 In the context of the prey-predator interactions, one can derive the encounter rate kernels for the various cases, to obtain the following relations:

\begin{gather}
4 \pi DR : \text{Random walk} \\
1.3 \gamma R^{3} : \text{$<$ Kolmogorov scale} \\
1.37 \pi R^{2}(\epsilon R)^{1/3} : \text{$>$ Kolmogorov scale} \\
4/3 \pi R^{3}f : \text{Stop-and-go motion} \\
\pi R^{2}u : \text{Swimming, sinking or feeding current Brownian diffusion} 
\end{gather}
 
The general framework of encounter rate calculations considers a spherical encounter zone with an effective size \cite{kiorboe2018book}, which has been recently extended to account for non-spherical morphologies \cite{andersen2019}. The initial models which focused on the encounter rates under diffusive and laminar flow fields, were extended for the turbulent eddies by Rothschild and Osborn \cite{rothschild1988}, wherein velocities of both agents (for instance, the prey and the predator) we included. The extended encounter rate model demonstrated that turbulence could increase the contact rates by 50\% or more, depending on the size and swimming rate \cite{lewis2000}. More recently, Arguedas–Leiva \textit{et al.}, have showed that the encounters between neutrally buoyant elongated cells are up to ten-fold higher relative to spherical cells, with further enhancement in encounter rates for those which sink, instead of being neutrally buoyant \cite{arguedas2022}.


\section{Gravitaxis in planktonic active matter}\label{gravitaxis}

Gravitaxis refers to the movement of organisms in response to the gravity vector. Most phytoplankton species exhibit gravitaxis, as a means to execute diel vertical migration. Historically, the movement of gravitactic species along the gravity vector was referred to as \textit{geotaxis} \cite{roberts1970,bean1984book,fenchel1984}, however in light of the generality of this tactic response to gravity forces - not only of the Earth but also to that due to other planetary bodies or artificial accelerations - the term \textit{gravitaxis} became more widely used \cite{Braun2018gravisensing}. Today, alongside \textit{gravitaxis}, the term \textit{gyrotaxis} is frequently used, particularly to appropriately capture and describe the interplay of gravity with fluid forces that are ubiquitous in the watery environments which the plankton inhabit \cite{kessler1985,kessler1985-2,kessler1986, pedley1990,pedley1992,jones1994}. The ability of organisms to swim against (negative gravitaxis) or along the gravity vector (positive gravitaxis) depends on the cells' developmental phase, physiological state, time of the day or season, and response or adaptation to exogeneous stressors \cite{Braun2018gravisensing-2, sengupta2017, carrara2021, sengupta2022}. In the following sections, we will, step-by-step, discuss how planktonic active matter perceive gravity forces, and leverage the gravity-flow interactions to navigate different fluid dynamic and ecological settings.

\subsection{Gravity-sensing mechanisms}
Planktonic microbes sense and respond to the gravitational forces and changes therein using a series of steps: perception, transduction, followed by signal amplification and response, emmploying different receptors capable of detecting the gravity signals either directly or indirectly. Organelles for direct sensing include heavy BaSO$_{4}$ crystals which function as statoliths, or SrSO$_{4}$ crystals which are also used in the statocyst-like organelles of the ciliates \textit{Loxodes} and \textit{Remanella} \cite{Braun2018gravisensing-2}. Such heavy organelles - found across diverse organisms spanning ciliates and algae - operate by directionally moving (sedimenting) within cells, thereby initiating a mechano-signal transduction chain that ultimately allows cells to distinguish between up versus down, or sense changes in their local accelerations \cite{limbach2005, strohm2012}. In some larger planktonic species lacking heavy statoliths, the entire cytoplasmic content of the cell can proxy as a gravity-sensing organelle, exerting pressure on the lower membrane, thereby activating the mechano-(gravi-)sensitive ion channels distributed in the cell membrane \cite{hader2017book}. 

The exact mechanism by which small phytoplankton perceive gravity forces and changes therein (in turn, changes in their orientation relative to gravity vector) remains unclear, and thus warrants further investigation. Large ($>75~\mu$m) unicellular protists like the \textit{Paramecium} and \textit{Tetrahymena} sense gravity by an active physiological mechanism through calcium or potassium mechanosensitive ion-channels \cite{hemmersbach1999,hemmersbach2002,richter2002}, which get activated due to the gravitational pressure of the cytoplasm on the lower membrane. In the flagellate \textit{Euglena gracilis}, typically 35–50 $\mu$m in size, mechanochemical changes in the membrane potential are involved in graviperception \cite{hader2001,richter2003}. For cells in the size range of the raphidophyte \textit{Heterosigma akashiwo} (10–15 $\mu$m, Figure \ref{fig:mechanics}), the gravitational force on the lower membrane can be approximated by: 

\begin{equation}\label{eqn:membrane_force}
F = \delta \rho V \text{g}
\end{equation}

where, $\delta \rho$ difference in the density of the cell and the surrounding fluid, $V$ is the cell volume, and \textbf{g} represents the acceleration due to gravity (or, any relevant acceleration, in general). For microscale ciliates, this yields a force of the order of tens of pN, while for the larger ciliates, e.g., \textit{Paramecium caudatum}, this results in a force over 100 pN \cite{hader2005book}.

The magnitude of the force yields work of the order of the thermal noise, assuming that the entire cytoplasmic material functions as a buoy. The work due to the gravitational force
on the lower membrane for 1 nm gating distance of the mechanosensitive ion channels can be estimated as $4 \cdot 10^{-22}$ J, while the thermal noise, $kT/2$, at room temperature (293 K) is
$\approx 2 \cdot 10^{-21}$ J, where $k$ is the Boltzmann constant \cite{sengupta2017}. This suggests the possibility of alternative mechanisms for gravity-sensing, particularly for microplankton which lack any other direct sensing mechanism. One such alternative could be the cross-talk between the ion-channels and the production of reactive nitrogen species (RNS), wherein the sensing may involve positive feedbacks between these two cellular networks \cite{carrara2021, besson-bard2008}.

\subsection{Biomechanics of gravitaxis}

\begin{figure}[htp]
	\begin{center}
		\includegraphics[height=6cm]{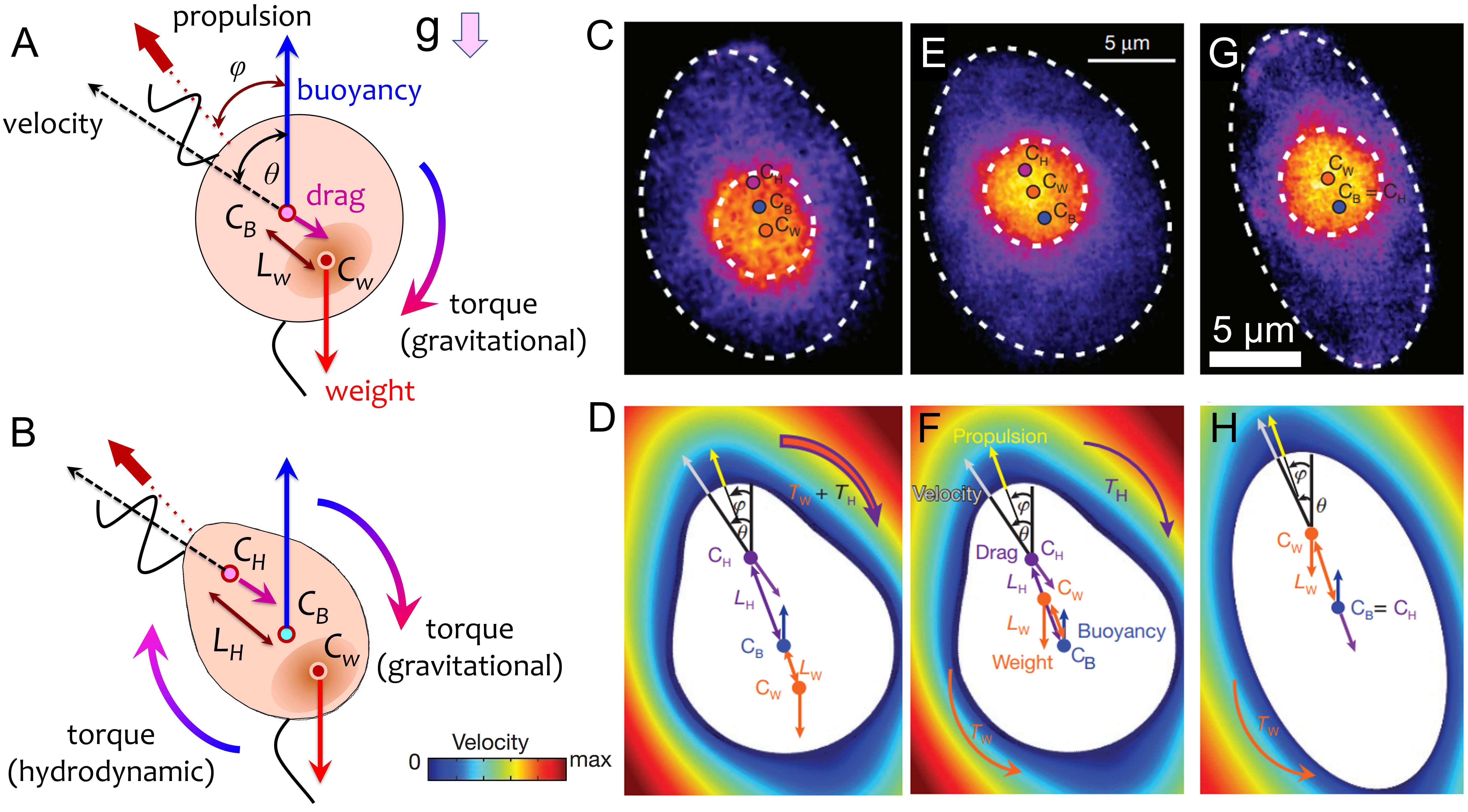}
	\caption{{\bf Cell mechanics of gravitaxis.} 
	{\bf a} Schematic of a symmetrically-shaped bottom-heavy microplankton swimming at low Reynolds number. The cell experiences a propulsion force due to the beating flagellum (or flagella/cilia), and body weight, both acting at the center of mass, $C_{W}$. The buoyancy force acts at the geometric center of the cell body (centroid), $C_{B}$, which is offset from the center of mass by a distance $L_{W}$. The buoyancy force (opposite to the gravity force) and the drag force (opposite to direction of swimming) act through $C_{B}$. When the cell is displaced by an angle $\theta$ from its equilibrium swimming direction, a stabilizing torque reorients the cell back to the equilibrium orientation (here, in the clockwise direction) due to the bottom heaviness ($C_{W}$ below $C_{B}$). {\bf b} For a cell with fore-aft asymmetry, two separate offset lengths emerge: $L_{W}$ (as above), and $L_{H}$, due to the offset between the centroid and the center of hydrodynamic stress $L_{H}$. For a symmetric cell shape, the $L_{W}$ and $L_{H}$ coincide. The two different offset lengths generate distinct reorientation torques about the centroid of the cell: gravitational ($T_{W}$) and hydrodynamic ($T_{H}$). The two torques can either reinforce each other (act in the same direction, in the case that the cell is bottom heavy), or counteract if the cell is top-heavy ($C_{W}$ above $C_{B}$). {\bf c} Epifluorescent image of bottom-heavy microplankton cell with fore-aft asymmetry, showing the $C_{W}$, $C_{B}$ and $C_{H}$, and the corresponding free-body diagram, shown in panel {\bf d}. The central orange hue represents the cell nucleus, the heaviest organelle within the cell body. Both $T_{W}$ and $T_{H}$ act to orient cells upwards. {\bf e, f} Top-heavy asymmetric cell swimming at an angle $\theta$, such that the rotation rate $\omega$ are set by the competition between the gravitational and hydrodynamic torques about $C_{B}$, causing cells to orient upwards. {\bf g, h} For the top-heavy symmetric cells, the $T_{H}$ vanishes, thus causing the cells to orient downwards. Panels {\bf c}-{\bf h} have been adapted from Ref. \cite{sengupta2017}.
	}\label{fig:mechanics}
	\end{center}
\end{figure}

 Assuming a body of revolution swimming in a fluid at a low Reynolds number, the translational and rotational equations of motion are decoupled \cite{roberts1970,roberts2002,roberts2006,roberts2010}, allowing us to write the following equations along the major-and the minor-axis (Figure \ref{fig:mechanics}, under force-free conditions:

\begin{equation}\label{eqn:PAM:symmetric}
\begin{gathered}
    P \sin \phi  = D \sin \theta \\
    P \cos \phi - D \cos \theta = (\rho_{cell} - \rho_{fluid})Vg
\end{gathered}
\end{equation}

where $P$ is the propulsion force originating due to the beating of the flagellum, acting along the long axis of the cell body, an angle $\phi$ to the gravity vector; $D$ is the drag force acting through the center of the hydrodynamic stress $C_{H}$, directed opposite to the cell swimming velocity, at an angle $\theta$ relative to the vertical. The volume of the cell, and the densities of the cell and the surrounding fluid are $V$, $\rho_{cell}$ and $\rho_{fluid}$ respectively. The drag force, $D$ on the moving body in a fluid with dynamic viscosity $\eta$ at a velocity $v$ depends on the angle $\alpha = \theta - \phi$ between the body axis and the direction of swimming; and can be can be broken down into two orthogonal components as: 

\begin{equation}\label{eqn:PAM:drag}
    D = D_{\parallel} \cos \alpha + D_{\perp} \sin \alpha
\end{equation}

where $D_{\parallel}$ and $D_{\perp}$ are the drag forces corresponding to motion along and perpendicular to the direction of the major axis of the body, respectively. 

Another independent set of equations arises due to the balance of the torques acting at the cell's center of buoyancy (i.e., the geometric center), $C_{B}$, giving: 

\begin{equation}\label{eqn:PAM:torque}
    T_{V} = T_{H} + T_{W}
\end{equation}

where $T_{H}$ is the torque generated by the drag force $D$, $T_{W}$ is the torque generated by the weight of the cell $W = V \rho_{cell}g$. On the other hand, the propulsion force $P$ generates no torque about $C_{B}$ as it passes through $C_{B}$. The net torque balance equation can be then written as:

\begin{equation}\label{eqn:PAM:torquebal}
   D \sin (\theta - \phi)L_{H} - W [\sin(\phi - \arctan (L_{Nb}/L_{Na}))] L_{W} = R \eta \omega
\end{equation}

Here, $\arctan (L_{Nb}/L_{Na})$ is the contribution to the gravitational torque coming from the offset $L_{Nb}$ of the nucleus within the equatorial plane; and $R \eta \omega$ is the net viscous torque, $R$ being the coefficient of resistance of the body to rotational motion, and $\omega$ is the rotation rate of the cell. 

The length-scale $L_{H}$ is the offset distance between the center of buoyancy $C_{B}$ and the hydrodynamic stress center $C_{H}$ (Figure \ref{fig:mechanics}\textbf{c-h}). The centre of hydrodynamic stress is the point at which the resultant of all viscous stresses exerted by the fluid on the cell (resulting from the combination of translational motion, reorientation and sedimentation) acts. The center of hydrodynamic stress for bodies with spherical or cylindrical symmetry lies along the axis of symmetry. In case of cells with fore-aft asymmetry, one can obtain the position of $C_{H}$ by numerically solving the Navier–Stokes equations around the cell body, taking into account the characteristic size and shape determined experimentally by quantitative image analysis. When torque-free condition (sum of all torques on the cell vanishes) are applied, one obtains the coordinates of $C_{H}$, determined by minimizing the surface integral of the cross-product between the stress force and the surface of the cell. 

\begin{figure}[htp]
	\begin{center}
		\includegraphics[height=10cm]{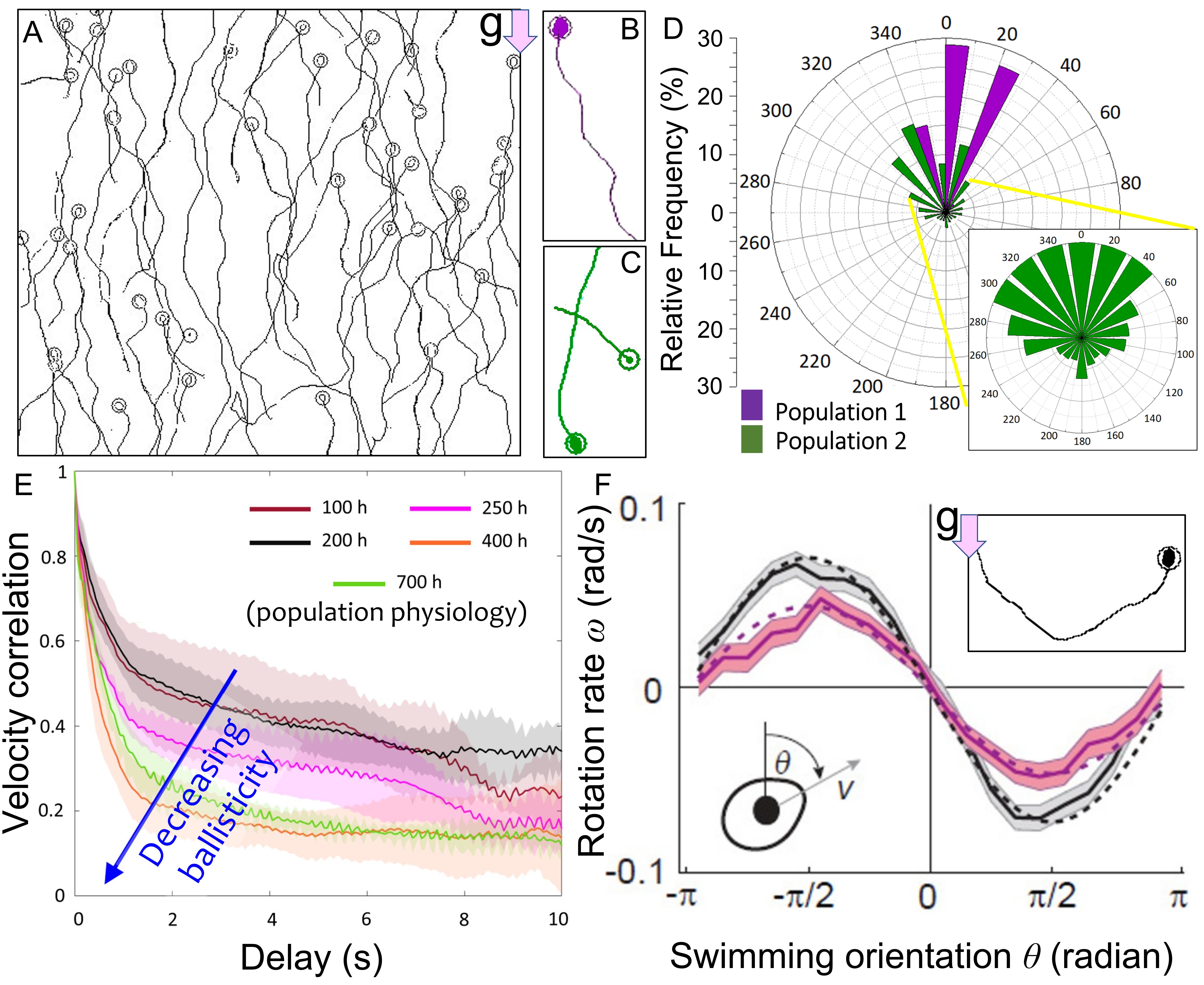}		
	        \caption{
	        {\bf Behaviour and stability of gravitactic swimming.} {\bf a} Trajectories of up-swimming gravitactic microplankton (swimming against the gravity direction). Trajecotry patterns (curvature and helicity) can be used to extract the orientational stability of swimming cells. {\bf b} Characteristic helical swimming trajectory of a gravitactic cell. {\bf c} Sample trajectories of down-swimming microplankton (swimming in the gravity direction). {\bf d} Windrose plot presents the strength and angular spread of the microplankton motility of physiologically distinct swimming populations: strongly ballistic swimming (purple) vs weakly ballistic swimming (green). Higher angular spread of the trajectories of the second population reduces the ballisticity (low swimming anisotropy). The zoomed-in view of the windrose center captures the angular distribution of the second population. Image adapated from \cite{sengupta2022}. {\bf e} Age-dependent swimming modulation in motile phytoplankton. A shift from ballistic ($t = 100$ h) to diffusive ($t = 700$ h) swimming is captured by plotting the velocity correlations over the delay time \cite{sengupta2022}. {\bf f} Rotation rate, $\omega$,as a function of the direction, $\theta$, of the instantaneous swimming velocity, $v$ relative to the vertical direction. The rotation rate of the cells, quantified by tracking them in different time intervals (short time shown in grey, while longer time is shown in magenta), averaged over all the cells as a function of $\theta$. The difference between the two curves denotes the presence of cells that reorient more rapidly and others that reorient more slowly. Dashed lines are sinusoidal fits to the experimental data, used to obtain the reorientation timescale $B$. Solid lines denote the arithmetic mean over all cell trajectories. The reorientation process is shown in the inset to the plot. Image adapted from Ref. \cite{sengupta2017}.
	        }\label{fig:stability} 
	\end{center}
\end{figure}

For a cell swimming with a speed $v$, we can numerically solve the system of above equations for the unknowns $P$, $\phi$ and $\omega$, yielding the rotation rate, $\omega(\theta)$, as a function of the swimming angle, $\theta$, with respect to the direction of the gravity force. The orientational stablity of the swimming cell can be extracted from the reorientation timescale $B$, by fitting a sinusoid in the $\omega(\theta)$ versus $\theta$ plots for individual cells. At the level of single cells, the reorientation trajectories corresponding to the different fitting parameters represent different reorientation timescales (Figure \ref{fig:stability}(\textbf{a,f}). At a population-scale, the overall anisotropy of the swimming velocity, represented in Figure \ref{fig:stability}(\textbf{b}, hints at the swimming stability of the individual cells: straight, anisotropic tranjectories and consequent distributions are correlated with high stability, $i.e.$ low reorientation timescales, and ballisticity of swimming (Figure \ref{fig:stability}(\textbf{e})) and vice versa.  

When a body of uniform density and arbitrary shape is immersed in a liquid the upthrust on the body acts through the centre of gravity, and there is no tendency for buoyancy forces to rotate the body. If there is a variation in density, however, a torque is experienced, the magnitude of which depends upon the density differences within the body. The reorientation torque produced by a cell of arbitrary symmetric volume $V$ and density $\rho_{cell}$ is:  

\begin{equation}\label{eqn:PAM:gentorque}
T_{W} = (\rho_{cell} - \rho_{fluid})VL_{W}g \sin \theta 
\end{equation}

Taking the opposing viscous torque into account, one can write: 
\begin{equation}\label{eqn:PAM:gentorque-2}
\begin{gathered}
R \eta \omega = (\rho_{cell} - \rho_{fluid})Vg \sin \theta L_{W} \\
\frac{d\theta}{dt} = [(\rho_{cell} - \rho_{fluid})VL_{W}g/R \eta ] \sin \theta
\end{gathered}
\end{equation}

Here, $(\rho_{cell} - \rho_{fluid})VL_{W}g/R \eta$ signifies the instantaneous rate of reorientation, and its inverse gives the reorientation timescale $B$.  

Changes in both the cell density and cell shape can impact orientational stability of swimming cells \cite{roberts2002,sengupta2017}. For instance, when a cell expels water or gas molecules by regulating intracellular vacuoles, both the density distribution within the cell and the cell shape undergo considerable change. The distribution of the body mass, for instance top-heavy versus bottom-heavy mass distributions can significantly impact motility properties, and can even alter swimming behaviours \cite{durham2009, durham2011, sengupta2022}. Given the resemblance of phytoplankton morphologies with prolate spheroids, they have been widely used to model the mechanics of gravitaxis. While the sedimentation rate of symmetrically-shaped prolate spheroids (which are denser than the surrounding fluid) can be estimated analytically \cite{happel1983book}, the calculations for distorted prolate spheroids with fore–aft asymmetry are more involved. One needs to account for the additional reorientation term, since asymmetric prolate sheroids rotate while sedimenting downwards at instantaneous rates given by the degree of asymmetry \cite{roberts1970, happel1983book, roberts2002, sengupta2017}. The sedimentation characteristics of an asymmetric cell body, with rotational symmetry, can be derived using the \textit{dumb-bell} model described by Happel and Brenner \cite{happel1983book}, which approximates the asymmetric fore-aft to be constructed equivalently out of two spherical blobs of different radii, connected by a light, rigid inextensible rod. For a single isolated spherical cell of radius $r$ and density $\rho_{cell}$, settling with a velocity $v$, the viscous drag (under low Reynolds number conditions) is balanced by the weight of the cell body, giving us:

\begin{equation}\label{eqn:PAM:sediment}
\begin{gathered}
\frac{4}{3} \pi r^{3} (\rho_{cell} - \rho_{fluid})g = 6 \pi \eta r v \\
v = \frac{2}{9} \frac{(\rho_{cell} - \rho_{fluid})g}{\eta} r^{2}
\end{gathered}
\end{equation}

Now, if the dumb-bell shaped cell made of the two spherical components (radii $r_{1}$ and $r_{2}$, each with density $\rho_{cell}$) interconnected by the light inextensible rod, is allowed to fall through the medium, one obtains:

\begin{equation}\label{eqn:PAM:shape}
v_{\parallel} = \frac{2}{9} \frac{(\rho_{cell} - \rho_{fluid})g}{\eta} \left(\frac{r_{1}^{3} + r_{3}^{3}}{r_{1}+ r_{2}}\right)
\end{equation}    

Here, $v_{\parallel}$ is the steady-state sedimentation speed the two-sphere dumb-bell with its long axis vertical (i.e. parallel to the gravity vector). Realistically, one needs to consider the hydrodynamic interaction between the interconnected spheres, which is ignored in the analysis of Happel and Brenner. If the total length of the model cell is $L$, with $e = r_{1}/L$ and $f = r_{2}/L$, the equation can be re-written as: 

\begin{equation}\label{eqn:PAM:shape2}
\begin{gathered}
v_{\parallel} = \frac{2}{9} \frac{(\rho_{cell} - \rho_{fluid})g}{\eta} L^{2} \left(\frac{e^{3} + f^{3}}{e+ f}\right) \\
v_{\parallel} = \frac{2}{9} \frac{(\rho_{cell} - \rho_{fluid})g}{\eta} L^{2} F_{\parallel}
\end{gathered}
\end{equation}    

where $F_{\parallel}$ is the dimensionless shape-factor independent of the physical size of the cell. 

By following the above steps, one can derive the corresponding shape-factor when the cell is sedimenting perpendicular to the gravity vector (note that the perpendicular orientation is unstable, with the instantaneous velocity varying from one end to the other). The instantaneous volocity of the dumb-bell is given by the average of the two spheres, as:

\begin{equation}\label{eqn:PAM:shape-perp}
\begin{gathered}
v_{\perp} = \frac{2}{9} \frac{(\rho_{cell} - \rho_{fluid})g}{\eta} L^{2} \left({e^{2} + f^{2}}\right) \\
v_{\perp} = \frac{2}{9} \frac{(\rho_{cell} - \rho_{fluid})g}{\eta} L^{2} F_{\perp}
\end{gathered}
\end{equation}    

Extending the above analysis, one can derive the orientational characteristics of the asymmetric dumb-bell, as a function of the angle between the long axis of body the gravity direction, $\theta$ : 
\begin{equation}\label{eqn:PAM:shape-orient}
\begin{gathered}
\frac{d \theta}{dt} = - \frac{1}{B} \sin \theta \\
\end{gathered}
\end{equation}   

where the maximum reorientation rate (at $\theta = \pi/2$) is given by $\frac{1}{B} = \frac{(\rho_{cell} - \rho_{fluid})}{\eta} g L F_{\theta}$. The shape-factor $F_{\theta}$ can be obtained experimentally.

Happel and Brenner \cite{happel1983book} have provided full analytical solutions for the hydrodynamic drag on a prolate spheroid of semi-major and semi-minor axes $a$ and $b$ moving both parallel and perpendicular to the long axis. The directional drag coefficients which an asymmetric prolate cell experiences is given by:

\begin{equation}\label{eqn:PAM:drag-directional}
\begin{gathered}
D_{\parallel} = 6 \pi \eta r_{eq}v F_{\parallel} \\
D_{\perp} = 6 \pi \eta r_{eq}v F_{\perp}
\end{gathered}
\end{equation}  
 where, $r_{eq}$ is the radius of a sphere with volume equal to the prolate spheroid; and the directional shape-factors defined as:

\begin{equation}\label{eqn:PAM:drag-directional-2}
\begin{gathered}
F_{\parallel} = \frac{1}{48 \phi} \left[ \frac{-2 \phi}{\phi^{2}-1} + \frac{2 \phi^{2}-1}{(\phi^{2}-1)^{3/2}} \ln \left(\frac{\phi + (\phi^{2}-1)^{0.5}}{\phi - (\phi^{2}-1)^{0.5}} \right) \right] \\
F_{\perp} = \frac{1}{48 \phi} \left[ \frac{\phi}{\phi^{2}-1} + \frac{2 \phi^{2}-3}{(\phi^{2}-1)^{3/2}} \ln \left(\phi + (\phi^{2}-1)^{0.5} \right) \right]
\end{gathered}
\end{equation}  

 where $\phi = a/b$, $a$ and $b$ being the semi-major and semi-minor axes respectively. The coefficient of resistance of a prolate spheroid to rotational motion is is given by \cite{koenig1975}:

 \begin{equation}\label{eqn:PAM:drag-rotational}
R (\phi) = 8 \pi r_{eq}^{3} \frac{2(\phi^{2}+1)(\phi^{2}-1)^{3/2}}{3 \phi  \left[ (2t^{2}-1) \ln \left(t + (t^2-1\right)^{0.5}) - t(t^2-1)^{0.5}\right]}
\end{equation}

\subsection{Quantifying cell morphology in experiments}

\begin{figure}
	\begin{center}
		\includegraphics[height=6cm]{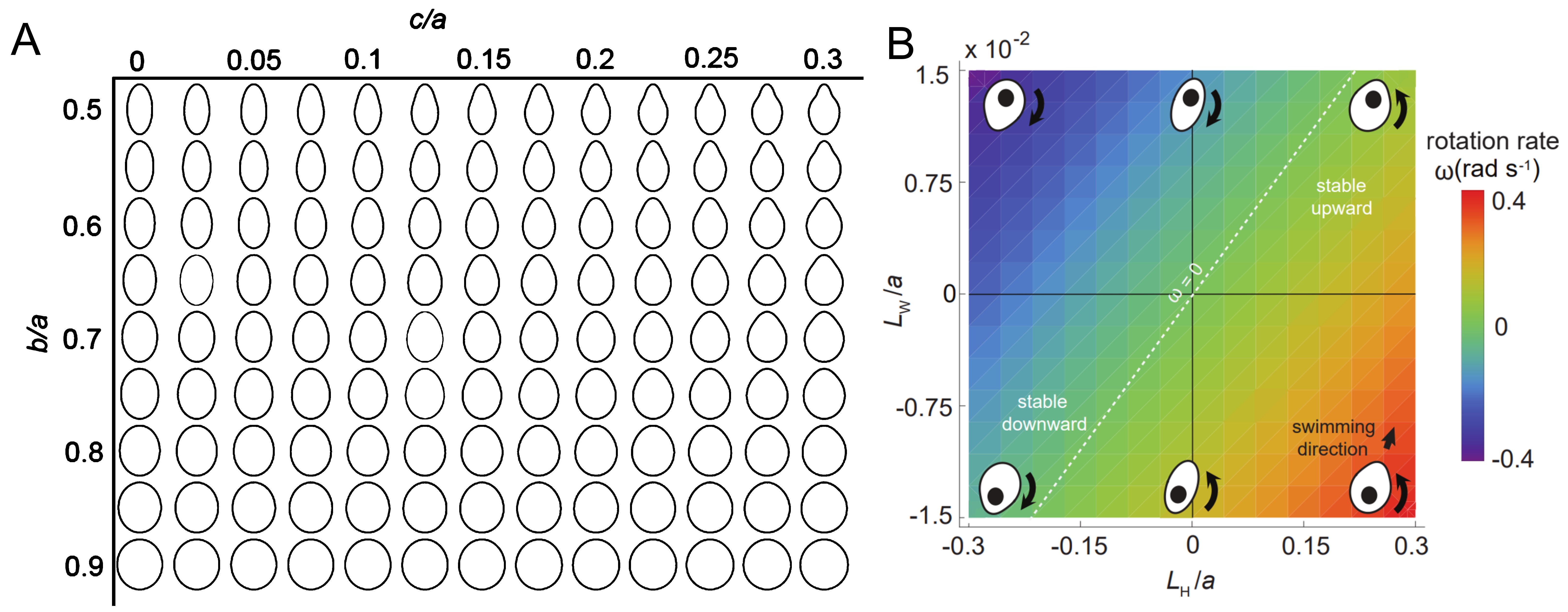}
	\caption{{\bf Gravitaxis of shape-shifting microplankton.} 
		   	{\bf a} The graph shows cell shape variation in terms of the degree of fore–aft asymmetry and minor/major axis ratio. The parameter $c$ denotes the degree of fore–aft asymmetry, $a$ is the semi-major axis, $b$ is the semi-minor axis. {\bf b} Two relevant physical features are presented two morphological length scales which determine the cell stability: the asymmetry in shape, quantified by $L_{H}/a$, and the mass distribution, quantified by $L_{W}/a$, where $a$ is the semi-major axis, $L_{H}$ quantifies the distance between the centre of buoyancy and the centre of hydrodynamic stress, and $L_{W}$ is the distance between the centre of buoyancy and the centre of mass. The colours denote the cell rotation rate $\omega$ following an orientational perturbation. $\omega > 0$ denotes negatively gravitactic cells (stable upward), while $\omega < 0$ denotes positively gravitactic cells (stable downward), and $\omega = 0$ (white dashed line) denotes neutrally stable cells. Sample asymmetry configurations corresponding to different locations on the regime diagram are illustrated by the schematics. Adapted from \cite{sengupta2017}.
	}\label{fig:fitting}
	\end{center}
\end{figure}

\begin{figure}
	\begin{center}
		\includegraphics[height=6cm]{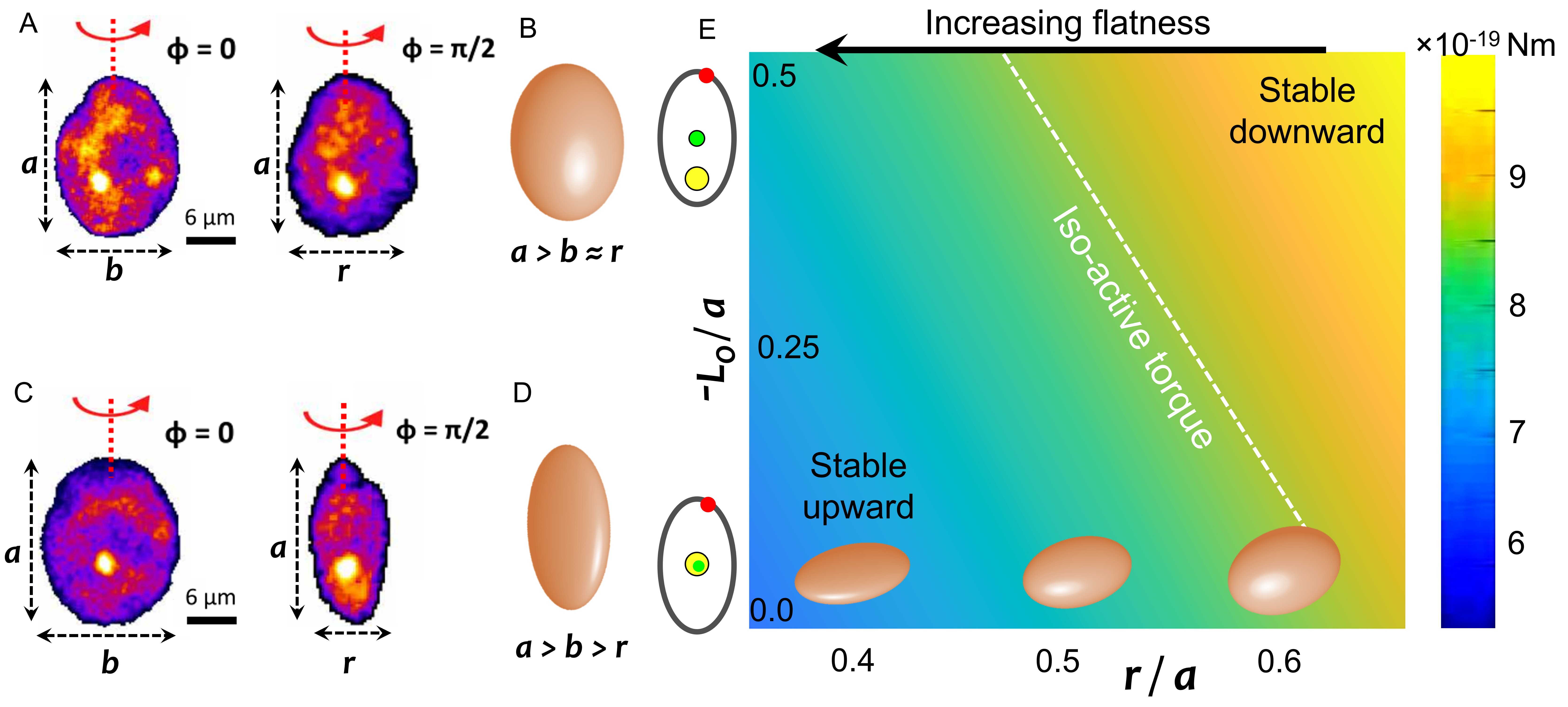}		
	        \caption{{\bf Role of rotational symmetry on gravitaxis.} {\bf a, b} Axisymmetric morphology results in difference in swimming stability relative to the cells which are flat-shaped (lacking axisymmetry, {\bf c, d}). {\bf e} The active torque (in Nm) required by a cell to reorient itself as an upward swimmer (negative gravitaxis) as a function of the cell flatness $r/a$, as a function of the effective organelle offset distance $L_{O}$ (for instance, lipid position from the geometric center) and the semi-major axis $a$. The dashed white line shows the iso-active torque. Depending on the microplankton species, the active torque requirement decreases as the cells become flatter. Adapted from \cite{sengupta2022}.
	    }\label{fig:rotational}
	\end{center}
\end{figure}

Swimming cells imaged in real time can allow us to acquire various projections of the cell morphology, and thereby reconstruct the three dimensional morphology. The imaged cells can be then analysed to extract the cell contours (from multiple projections), followed by appropriate curve-fitting of the extracted contours (Figure \ref{fig:fitting}). This technique has been fairly successful in quantifying the cell morphology of microplankon, starting with the pioneering works of Roberts \textit{et al.}, \cite{roberts2002,roberts2010}, and more recently by Sengupta \textit{et al.}, where changes in rotational symmetry of the cell morphology have been experimentally quantified, thus allowing a complete analysis of swimming stability based on the three-dimensional mrophological parameters (Figure \ref{fig:rotational}) \cite{sengupta2022}. A three-parameter equation has been found to reliably capture the projected morphological features of microplankton:

 \begin{equation}\label{eqn:PAM:shape5}
S (\gamma, \psi) = \frac{ab}{(a^{2} \sin^2 \psi + b^{2} \cos^{2} \psi)^{0.5}} + c \cos \psi
\end{equation} 

where where the first term on the right describes an ellipse with semi-major and semi-minor axes of lengths $a$ and $b$ respectively, and the second term confers a degree of fore–aft asymmetry specified by the length $c$. The angles $\gamma$ and $\psi$ are the polar and azimuthal angles measured from the major axis, and $S(\gamma, \psi)$ represents the distance of a point on the surface from the origin with a polar angle $\gamma$ and azimuthal angle $\psi$. Varying the relative ratio of $a/b$ alters the eccentricity of the symmetric ellipsoid of revolution, whereas by altering the relative ratio $c/a$ changes the fore-aft asymmetry of the cell morphology. Thus, plotting the values of $b/a$ and $c/a$ give the family of curves which varying eccentricities and fore-aft asymmetries as shown in Figure \ref{fig:rotational}\textbf{a}.

It is imperative to note here, that the above analyses have been carried out assuming the cell body to be rotationally symmetric, \textit{i.e.}, the cross-sectional plane normal to the long axis of the cell body has a circular shape (for instance, a pear or an egg). However, many planktonic species lack rotational symmetry, or depending on their physiological status, develop \textit{platelet} shape morphologies with certain degree of flatness  \cite{sengupta2022}. For such cases one has to account for the lack of rotational symmetry while calculating the gravitactic stability, since the translational and rotational viscous torques are direction dependent. The drag force $D$ on an arbitrary ellipsoid with semi-axes $a$, $b$ and $r$ (Figure \ref{fig:rotational}), swimming within a fluid with speed $U$ along the long axis ($a$-direction) is :

\begin{equation}\label{eqn:PAM:drag-general}
\begin{gathered}
    \frac{D}{\pi \eta U} =  \frac{16}{\delta + a^{2}\chi_{a}} \\
    \delta = \int_{0}^{\infty} \left[(a^{2} + x')(b^{2} + x')(r^{2} + x')\right]^{-0.5} \,dx' \\
    \chi_{a, b, r} = \int_{0}^{\infty} \left[((a,b,r)^{2} + x')[(a^{2} + x')(b^{2} + x')(r^{2} + x')]^{0.5} \right]^{-1} \,dx'
\end{gathered}    
\end{equation}

where $a, b, r$ denote the individual relationships for $\chi_{a} \chi_{b}, \chi_{r}$ respectively. The resistive torque applied by the surrounding fluid due to rotation of the solid body along the long axis with an angular speed $\omega$ is given by: 

\begin{equation}\label{eqn:PAM:drag-asymm}
    \tilde{R} = C_{0}\frac{b^{2} + r^{2}}{b^{2}\chi_{b}+r^{2}\chi_{r}}
\end{equation}

where, $C_{0}$ is a constant pre-factor. The integrals can be calculated numerically, noting that the integral is sensitive to initial discretisation of $x^{'}$. Taking the symmetric geometry as a validation case, the discretisation can be accurately estimated. Furthermore, the accuracy of the technique can be verified by comparing the resistive viscous torque for the asymmetric case with that of the symmetric case: the two should equate when the semi-minor axes tend to similar values, i.e., when $b \to r$ (Figure \ref{fig:rotational}\textbf{e}) .

\subsection{Influence of intracellular organelles on gravitactic stability}

Planktonic cells contain a range of intracelllar organelles within their cell body, including the nucleus (the heaviest organelle within the cell), statoliths (miniscule biomineralized particles which enable gravity and pressure perception), biomineralized globules, chloroplasts, gas vacoules, lipid bodies and carbohydrate or starch reserves. Depending on the growth conditions and physiological constraints planktonic species experience, intracellular organelles can provide direct biomechical cues to actively modulate bouyancy, swimming speed and orientational stability of gravitactic species. Furthermore, the orientational stability of the cells could be fine-tuned through dynamic variation of the organelle size and their relative intracellular positioning within the cell cytoplasm. To understand the impact of the organelles on the swimming stability, one can approach the analysis in the framework of an \textit{effective} size, density and position (center of gravity), and how these impact the corresponding cellular parameters. The geometric and the hydrodynamic centers of the cell remain unaffected as long as the overall morphology of the cell remains intact.

Consider a cell with a generic morphology of volume $V_{C}$, which along with its nucleus (volume $V_{N}$, density, $\rho_{N}$), also contains dynamic intracellular organelles (total volume $V_{O}$, density, $\rho_{O}$). In case the effective center of gravity of the organelles, $C_{O}$, is located away from the long-axis of the cell body, it will generate a rotational moment about the geometric center of the cell (center of buoyancy, $C_{B}$), yielding the following set of equations: 

\begin{equation}
\begin{gathered}\label{eqn:PAM:organelles-force}
    P \sin \phi  = D \sin \theta \\
    P \cos \phi - D \cos \theta = (\rho_{cyto} - \rho_{fluid})V_{C}g + (\rho_{N} - \rho_{cyt})V_{N}g + (\rho_{O} - \rho_{cyt})V_{O}g
   \end{gathered}
\end{equation}

where, $\rho_{cyto}$ and $\rho_{fluid}$ are the densities of the cell cytoplasm (the intracellular material); and $\phi$ and $\theta$ are the respectively the angles of the cell propulsion ($P$) and the swimming velocity ($v$) to the vertical gravity direction (realistically, a cell does not swim exactly in the direction of propulsion). Balancing the torques, one obtains:

\begin{equation}
\begin{gathered}\label{eqn:PAM:organelles-torque}
       D \sin (\theta - \phi)L_{H} - W_{N}L_{N} \sin (\phi_{N}) - W_{O}L_{O} \sin (\phi - \phi_{O}) = R \eta \omega 
\end{gathered}
\end{equation}

Here, $L$ denote the off-set distance from center of buoyancy ($C_{B}$), $W$ the weights, while the subscripts $N$, $O$, and $H$ refer to the center of gravity due to the nucleus, organelles and the hydrodynamic center respectively. $\eta$ and $\phi_{O}$ respectively refer to the fluid viscosity and the angle between the cell's long-axis and the line joining $C_{B}$ and $C_{O}$. The above general system of equations can be extended to include the contributions of other intracellular organelles, for instance, if the cell has $n$ distinct organelles scattered within the cell at $\left(L_{O,i}, \phi_{O, i}\right)_{i=1,n}$, one arrives at:

\begin{equation}
\begin{gathered}\label{eqn:PAM:organelles-torque-2}
    P \sin \phi  = D \sin \theta \\
    P \cos \phi - D \cos \theta = (\rho_{cyto} - \rho_{fluid})V_{C}g + (\rho_{N} - \rho_{cyt})V_{N}g + \sum (\rho_{O, i} - \rho_{cyt})V_{O, i}g \\
    D \sin (\theta - \phi)L_{H} - W_{N}L_{N}\sin (\phi_{N}) - \sum \left[W_{O,i}L_{O,i}\sin (\phi - \phi_{O,i})\right] = R \eta \omega 
 \end{gathered}
\end{equation}

\subsection{Gyrotaxis: An interplay of gravitactic and viscous torques}

Gravitactic phytoplankton often encounter diverse hydrodynamic settings, including shear flows and turbulent eddies \cite{barry2015, durham2009,durham2011,durham2013,lillo2014,lovecchio2017,lovecchio2019}. In this section, we outline the interaction between gravity and hydrodynamic forces acting on motiles cells. When associated with flowing liquids, gravitactic cells additionally experience viscous torque: for a spherically symmetric (radius, $a$), bottom-heavy cell swimming with a constant speed $v_{C}$, the net torque acting on the cell is given by:

\begin{equation}
\begin{gathered}\label{eqn:PAM:gyrotaxis}
    T_{gyro} = 8 \pi \mu a^{3} \left[ \frac{\nabla \times u}{2} - \omega \right] + ML_{W} \times g
 \end{gathered}
\end{equation}

where \textbf{u} and $\mu$ are velocity and dynamic viscosity of the surrounding fluid, $\nabla \times u$ is the vorticity, and $\omega$ is the angular speed of the spherical cell. For micron-sized or smaller swimming cells, the Stokes flow regime holds good \cite{kiorboe2018book}, and under weak vorticity conditions, we can derive the characteristic scales of a gyrotactic swimmer: 

\begin{equation}
\begin{gathered}\label{eqn:PAM:gyrotaxis-equil}
    \beta = \frac{4 \pi \mu a^{3} v_{C}}{mgL_{W}}  = \frac{3 \mu}{\rho_{C}gL_{W}} v_{C} = B_{gyro}v_{C}
\end{gathered}
\end{equation}

where $\beta$ and $B_{gyro}$ are the gyrotactic length and timescales respectively. The equilibrium swimming direction, $\theta$, is given by:

\begin{equation}
\begin{gathered}\label{eqn:PAM:gyrotaxis-equil2}
    \sin \theta = \frac{3 \mu}{\rho_{C}gL_{W}} \times \frac{du}{dr} = B_{gyro}S
 \end{gathered}
\end{equation}

 which depends on $S = \frac{du}{dr}$, the gradient of flow velocity around the cell.
 
 Gradients in the surrounding flow flow may disrupt gravitactic migration of phytoplankton species, trapping them within layers, with overall, different thicknesses depending on the relative strength of the viscous and gravitactic torques \cite{durham2009}. Gyrotactic trapping occurs for $S > 1/B_{gyro}$, $i.e.$, when the gravitational torque upwards is outweighed by the reorienting hydrodynamic torque, driving the cells to tumble and accumulate over time. On the other hand, for $S < 1/B_{gyro}$, the swimming cells evade the \textit{gyrotactic trap}, and continue swimming along their gravity-mediated trajectories. The critical gradient of fluid velocity, $S_{cr}$, which induces gyrotactic trapping is given by:

\begin{equation}
\begin{gathered}\label{eqn:PAM:gyrotaxis-equil3}
    S_{cr} = 1/B_{gyro} = \left(\frac{3 \mu}{\rho_{C}gL_{W}}\right)^{-1}
\end{gathered}
\end{equation}

A generalization of the above can be done by including the effect of steady vortical flows: this leads to the formation of clustered aggregations of microorganisms \cite{durham2011}, mediated by the swimming speed and the orientational stability against the overturning due to the vorticity. The following generalized equations can be written:

\begin{equation}
\begin{gathered}\label{eqn:PAM:gyrotaxis-vortocal}
   \frac{d\mathbf{p}}{dt^*} = \frac{1}{2B}\left[\mathbf{k - (k\cdot p)p} \right] + \frac{1}{2}\omega^* \times \mathbf{p} + \alpha\mathbf{p} \cdot \mathbf{E^* \cdot \left[I - pp\right]}  
\end{gathered}
\end{equation}

where the starred quantities indicate the following dimensional variables: $\omega^*$ is the fluid vorticity, $\mathbf{E^*}$ is the rate of strain tensor, $\mathbf{I}$ is the identity matrix, $t^*$ is time, $B$ the characteristic reorientation timescale (to return to equilibrium orientation $\mathbf{k}$ if  $\omega^* = 0$, and $\alpha = (\gamma^2 -1)/(\gamma^2 + 1)$, $\gamma$ being the ratio of the major and minor axes of the cell body (assuming prolate ellipsoid geometry). 

The above equation applies to organisms much smaller in size than the scale of ambient velocity gradients (cells can be modeled as point particles). In experiments, swimming cells can form clusters within timescales of few minutes to hours, suggesting that vortical flow fields could be leveraged to separate planktonic phenotypes with distinct swimming stabilities. Note that for cells with no preferred swimming direction ($1/B = 0$), the Jeffery orbits are recovered \cite{guasto2012}. Using the Taylor-Green Vortex flow formalism, one can write the non-dimensional equations of motion as: 

\begin{equation}
\begin{gathered}\label{eqn:PAM:gyrotaxis-vortical-dimensionless}
   \frac{d\mathbf{p}}{dt} = \frac{1}{2\Psi}\left[\mathbf{k - (k\cdot p)p} \right] + \frac{1}{2}\omega (\mathbf{X}) \times \mathbf{p} + \alpha\mathbf{p} \cdot \mathbf{E(\mathbf{X}) \cdot \left[I - pp\right]} \\
   \frac{d\mathbf{X}}{dt} = \Phi \mathbf{p} + \mathbf{u(X)}
\end{gathered}
\end{equation}

where, $\mathbf{X} = \left[x, y, z \right]$, $\Psi = B \omega_{0}$, $\Phi = mV_{C}/\omega_{0}$, and the time non-dimensionalized by $\omega_{0}^{-1}$. The parameters $\Phi$ and $\Psi$ respectively measure the swimming speed relative to the flow field and the orientational stability of the swimming cell. The critical vorticity at which a cell is overturned by the vorticity is given by: 

\begin{equation}
\begin{gathered}\label{eqn:PAM:vortical-critical}
   \omega \Psi > 1 
\end{gathered}
\end{equation}

Various combinations of the $\Phi$ and $\Psi$ values highlight the regimes of phytoplankton aggregation in vortical flows, which arise due to the vorticity-motility interplay. 

Additional factors can contribute to the flow-induced rotational dynamics of motile species, and their emergent spatio-temporal distributions. For instance, the flagellar length can impact the cell geometry (rendering the cell more elongated than its body alone), or by modulating the drag associated with flagellar movement \cite{barry2015}. The impacts on the rotational dynamics can be further affected by active changes in behavior, morphology or cell's body mass distribution, emerging as a response to external cues \cite{Braun2018gravisensing-2,sengupta2017,carrara2021,sengupta2022}. Finally, certain phytoplankton species may actively resist the viscous torques, eliciting anomalous orientational distributions, in contrast to the Jeffery orbit predictions \cite{chengala2013,rafai2010,leahy2013}.

\subsection{Phytoplankton swimming in turbulent eddies}

Turbulence, along with light and nutrients, is known to be a primary determinant of plankton fitness and succession \cite{margalef1978mandala}. The hydrodynamic fluctuations associated with turbulent settings, frequently intermittent and localized within relatively thin horizontal layers \cite{thorpe2007,doubell2014,lozovatsky2015}, impact both swimming speed and trajectory (orientation). Recent studies suggest that phytoplankton may utilize physiological markers like ROS production and accumulation within cell bodies over time to sense turbulent environments, and adapt their migratory behaviour accordingly \cite{sengupta2017, carrara2021}. As shown in Figure \ref{fig:MD:ros}, phytoplankton species can actively switch their swimming direction under strong turbulent cues, thereby minimizing risks of any mechanical damage. Following the discussions in the previous section, the characteristic timescale of gravitactic reorientation depends on $B$ (the gyrotactic timescale) and $t_{\eta}$ (timescale corresponding to the velocity gradient). For the limiting case of weak (or strong) turbulence, $B \ll t_{\eta}$ (or $B \gg t_{\eta}$, the effect of turbulence is a perturbation from the stable swimming orientation (or stochastic orientations, captured by the orientational probability distribution of the swimmers). According to Lewis \cite{lewis2003, qui2022}, the orientation distribution of spherical gyrotactic swimmers in an isotropic turbulence is:

\begin{figure}
\begin{center}
		\includegraphics[height=6cm]{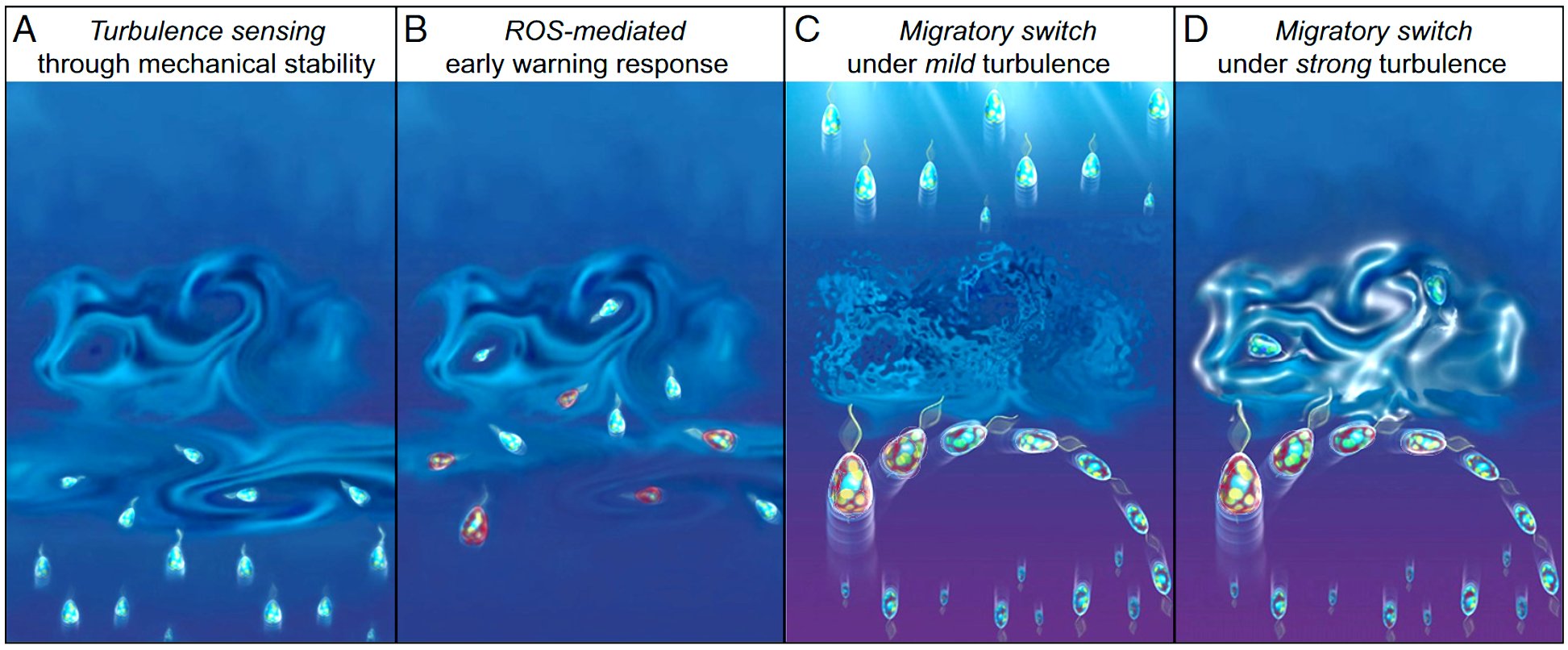}		
	        \caption{
	        {\bf Gravitaxis-mediated early warning mechanisms in phytoplankton species.}
	   	A ROS-mediated early-warning system enables microplankton populations to evade potential biomechanical damages (for instance shearing of flagella or even death) due to turbulence. {\bf a} Turbulent patches can disrupt cells’ migratory patterns, specifically when a population migrates vertically against the gravity direction. Cells can sense the intensity of a microscale turbulent eddy through their mechanical stability. Less stable cells are more easily thrown off balance and tumble under the effect of turbulence. {\bf b} Cells achieve the sensing attribute by integrating the ROS signals over multiple tumbling events, within timescales of tens of seconds. {\bf c} A negatively gravitactic population splits into two sub-populations: one that continues to migrate upward and one that switches to downward migration to avoid the turbulence. Cells, shown in red, with higher levels of ROS switch swimming direction from up to down, while others contnue swimming upward. {\bf d} Turbulence avoidance by adaptive swimming minimize biomechanical damages to the downward-swimming cells, however they show a reduced growth because of elevated ROS levels. However once the turbulence ceases, the down-swimming population scavenges excessive ROS and thereby recovers fitness. Adapted from \cite{carrara2021}.
	    }\label{fig:MD:ros}
	\end{center}
\end{figure}

\begin{equation}
\begin{gathered}\label{eqn:PAM:vortical-critical2}
   G(\theta) = \frac{\beta e^{\beta \cos \theta}}{2\sinh \beta} \text{;}
   ~\beta = \frac{B^{-1}}{2D_{eff}}
\end{gathered}
\end{equation}

where, $G$ is the probability distribution of $\theta$, and $D_{eff}$ is the effective diffusivity. Interstingly, the $G (\theta)$ distribution can be obtained also by deriving the Fokker-Planck relation from the constituitive equations, in the limit of turbulence represented by Gaussian noise. For this relation to be valid, along the trajectory of a swimmer, the correlation time of the turbulent velocity gradients is much less than the timescale of the change in the orientation. Effectively, this leads to $\Psi \gg 1$ or $\Psi < 1$ and $\Phi \Psi \gg 1$. The first condition corresponds to extremely weak gyrotaxis, while the second corresponds to the swimmers moving through the smallest flow scale in a short time so that the velocity gradients decorrelate quickly.

In the vicinity of turbulent eddies, motile phytoplankton experience centrifugal acceleration that tends to align the swimming direction toward the core of the eddy structure. This drives accumulation of active swimmers at the vortical core; in contrast, passive particles remain randomly distributed in the eddy environment as they get advected and dispersed, following the fluid streamlines \cite{durham2013,lillo2014}. Active planktonic swimmers are able to cross-over the streamlines, generating patches due to the interplay of the flow, swimming, gravity, and shape effects. The results have been subsequently confirmed by numerical studies, reporting the formation of patchiness across different scales \cite{breier2018patch,zhan2014}. Additionally, studies on the effect of cell shape on plankton patchiness \cite{borgnino2018shape,lovecchio2019} have revealed spatial distribution of elongated swimmers, as for the spherical swimmers, is fractal at small scales across a wide range of parameters. However, the concentration of cells in a cluster was found to depend on the morphology under similar swimming parameters. The variability in cell morphology, in combination with the swimming speed and orientational stability, allows a population of elongated swimmers ($e.g.$, in chain-forming motile species) to span both the downwelling and upwelling flow regions of a vortex\cite{gustavsson2016}. Other relevant factors including the shape of the swimmer, inertial effects and the nature of the turbulent environment (bulk turbulence versus surface turbulence) can impact gyrotactic clustering of planktonic active matter in natural settings \cite{qui2022,qiu2022-2}. The turbulence-induced patchiness is commonly quantified using the fractal dimension of the patches, $D$, calculated from the radial distribution function:

\begin{equation}
\begin{gathered}\label{eqn:PAM:vortical-critical3}
   g(r) = \frac{dN_{r}}{N(N-1)\text{d}r} \\
   g(r) \sim r^{D}
\end{gathered}
\end{equation}

where, $N$ is the total number of swimmers counted in the calculation of $g$, and D is defined as the exponent of $g(r)$ at small $r$. The value of $D$ indicates the spatial dimension when the swimmers are randomly distributed; and under clustered configurations, $D$ has smaller values. As an alternative metric to the fractal clustering, Voronoï tessellation has been employed to measure clustering \cite{liu2022}. The Voronoï tessellation splits the spatial domain into multiple polyhedrons of different volumes, the distribution of which provides information about the emergent patchiness. Here, a small polyhedron volume indicates that swimmers are in a local cluster, and vice versa. 

Alongside biophysical consequences, turbulence can drive physiological changes, including differential growth rates and photophysiology \cite{carrara2021}. Theoretically, for sufficiently large cells (equivalent radius greater $\sim$ 60–100 $\mu$m), turbulence may enhance mass transport and promote growth of cells under nutrient-limited settings \cite{guasto2012,kiorboe2018book}. In particular, for typical dissipation rates of turbulent kinetic energy, theoretical analysis indicates that natural levels of turbulence affect uptake by large microorganisms \cite{lazier1989,karpboss1996}, for instance, a cell with an equivalent radius $r = 50 ~\mu$m absorbing small molecules under strong turbulence ($\epsilon = 10^{-6}$ W/Kg) gains 18-32\% in nutrient uptake. While weak to moderate turbulence can enhance uptake and promote phytoplankton growth, strong turbulent cues may have deleterious effects on motile species, including physiological impairment, physical/biomechanical damage, and behavioural modification \cite{sengupta2017}. Biomechanical impacts of turbulence have been reported in \textit{Alexandrium minutum} which showed significant drop (by 50\%) in swimming velocity \cite{chen1998}, loss of flagella and swimming ability in \textit{Gonyaulax polyedra} \cite{thomas1990}, and disruption of the vertical migration by trapping \cite{durham2013}. On the physiological front, turbulence reduced the growth rates of certain species and, exposure over prolonged duration resulted in cell mortality, cellular disintegration, disruption of intrinsic cellular clock and biochemical cycles, and perturbed microtubule assemblage with potential effects on chromosome separation during cell division \cite{white1976,pollingher1981,berdalet1993,berdalet1992}. 

\subsection{Bioconvection}

\begin{figure}
	\begin{center}
		\includegraphics[height=6.5cm]{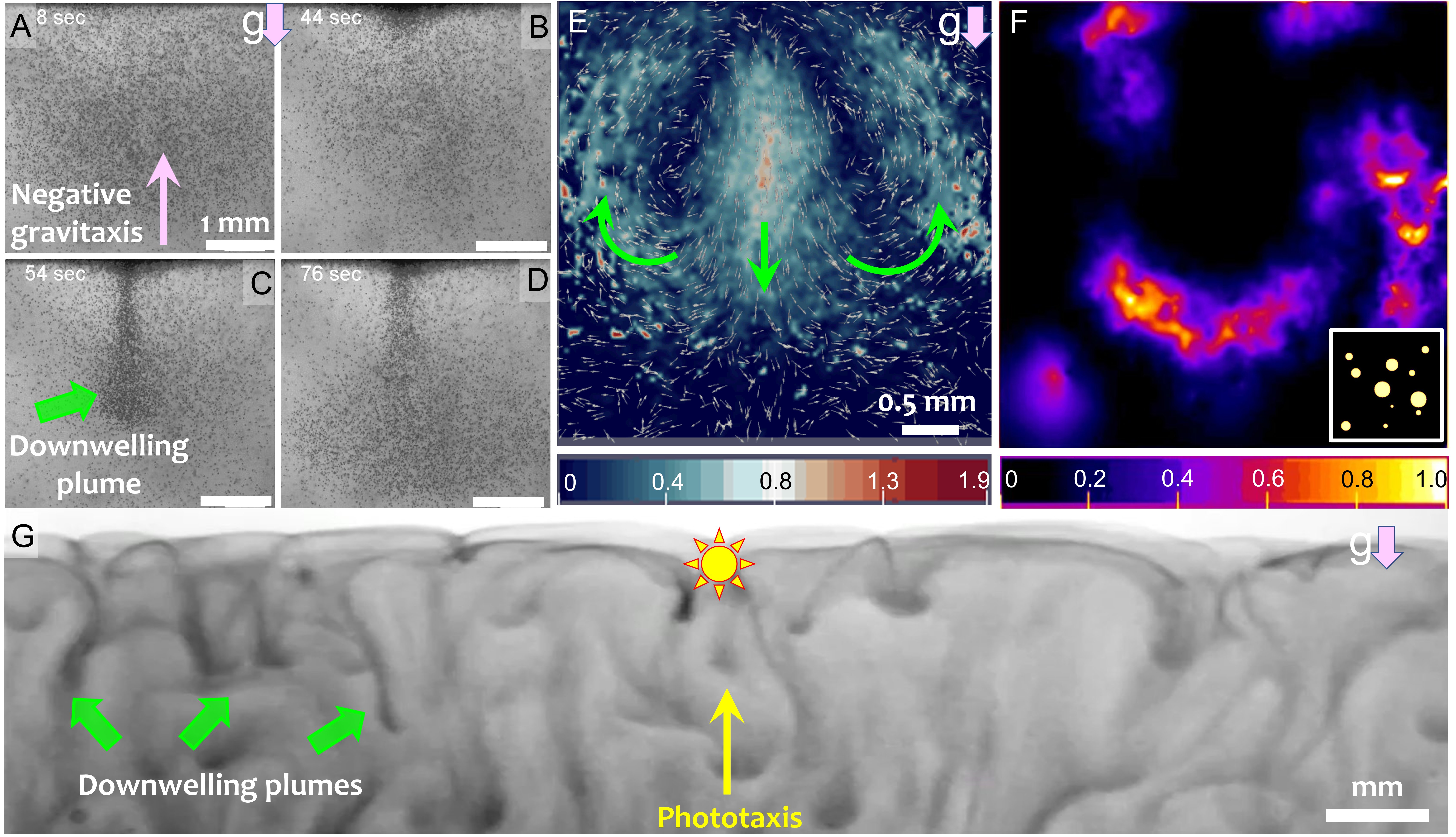}
	\caption{{\bf Bioconvection of gravitactic microplankton.} 
	{\bf a-d} Evolution of bioconvective plume due to negatively gravitactic spcecies \textit{Heterosigma akashiwo} confined within a millifluidic chamber. {\bf e} Analyzed image showing regions of high (red-brown) and low (blue-black) collective motion within a bioconvective plume. {\bf f} The plumes generate active motion of the surrounding liquid driving enhanced molecular transport. The  simulation snapshot shown here is based on experimental data, using computational fluid dynamics. The inset show an initial patch of high concentration of molecules. {\bf g} Similar bioconvective patterns can be see in diverse microorganisms, including phototactic bacteria \textit{Chromatium okenii} (image adapated from Ref. \cite{pfenning1960}) which swim toward sunlight and accumulates near the surface of lakes, before the onset of bioconvection. Such emergent flows in natural ecosystems have been implicated in distributing critical nutrients required by the microbes in their natural habitats. 
	}\label{fig:bioconv}
	\end{center}
\end{figure}

Bioconvection, the collective phenomena resulting in self-organized structures and emergent flow patterns in concentrated suspensions of swimming cells (Figure \ref{fig:bioconv}), has been known since the 19$^{\text{th}}$ century. The microbe-induced hydrodynamic instabilities and patterns arise due to the coupling between cell swimming; the biophysical properties of the cells such as the body mass distribution and density relative to the surrounding medium; and the ambient fluid dynamic settings \cite{bees2020}. Initially described as the “\textit{local movement of plant cells and their currents}” and the “\textit{emulsion figures and aggregation of spores in water}” by Nägeli (1860) and Sachs (1876) \cite{wager1911}, it was Wager’s comprehensive report in 1911 \cite{wager1911} that kickstarted the field of bioconvection. The study spanned host of species swimming under diverse conditions, over diurnal to seasonal timescales. Within 50 years of Wager’s seminal work, the term “\textit{bioconvection}” was first used by Platt \cite{platt1961} to describe patterns emerging in free-swimming organisms. The field of bioconvection saw steady growth during the last century, and continues expanding to date. 

Bioconvection initiates when negatively gravitactic microbes under confinement--either physical or established by competing gradients \cite{sommer2017}--swim and accumulate, thereby enhancing the local density of a region relative to the surrounding fluid. The difference in density triggers self-organized downwelling plumes packed with cells (Figure \ref{fig:bioconv} \textbf{a-d}), which are flanked by upwelling regions of up-swimming cells at relatively lower concentrations. Such local accumulation of cells, alternatively, can be driven by combinations of competing gradients (Figure \ref{fig:bioconv} \textbf{g}) including of light, chemicals or gases \cite{plesset1974,vincent1996,williams2011,sommer2017,prakash2021}. Plethora of swimming microorganisms, including bacteria, algae, and protozoa, have been observed to self-organize into bioncovective patterns at high concentrations. Bioconvection can emerge and persist purely due to mechanical reasons, for instance due to gyrotaxis \cite{bees2020,pedley-hill1988}. Non-uniform mass distribution within the cell bodies generate viscous and gravitational torques, which under conditions of bottom-heaviness (i.e., the bottom of the cell is heavier than the top half of the cell) and/or asymmetry in the cell shape, bestows upon cells the mechanical stability to swim upward against the gravity \cite{sengupta2017}. While bulk of the early studies have investigated bioconvection in the context of gyrotaxis, photosynthetic microorganisms–under natural settings–migrate under biophysical constraints posed by multiple confounding external cues (competing or reinforcing), including physiologically relevant light gradients (phototaxis), gravity forces (gravitaxis), and concentration of chemicals or gases (chemotaxis or aerotaxis) \cite{bees2020}. Depending on the species, confinement and strength of the gradients, bioconvective patterns can develop in timescales of minutes to hours; and yield characteristic length scales few orders of magnitude larger than the size of individual organisms. Qualitative similarities, particularly in regards to the collective vortex behavior and stigmergy, are observed between microbial bioconvective patterns and those generated by larger organisms including zooplankton, insects and fish \cite{bees2014maths,delcourt2016,houghton2018}.

The emergent macroscale collective phenomena can be described using a continuum model, accounting for the microscale behavior of individual swimmers. The model, based on the incompressible Navier-Stokes equations, includes the balance of the momentum, mass, cell numbers, and a description of the flow-dependent swimming stability (cell orientation moments). Generally, the Reynolds number for individual cells is much smaller ($\sim 10^{-3}$ for \textit{C. augustae}) than the characteristic Reynolds number of bioconvection ($\sim$10). The difference in the cell and fluid densities, $\rho$, at a cell concentration $n(x,t)$ and the mean volume $V_{mean}$, is incorporated with a negative buoyancy term \cite{childress1975,pedley1990} (subject to a Boussinesq approximation; in general, individual cells are a slightly heavier than the surrounding fluid, $\sim 5-10$\%). Thus, we obtain:

\begin{equation}
\begin{gathered}\label{eqn:PAM:vortical-critical4}
   \rho \frac{D\mathbf{u}}{Dt} = - \nabla p_{e} + nV_{mean} \Delta \rho \mathbf{g} + \nabla \cdot \Sigma \\
   \nabla \cdot \mathbf{u} = 0
  \end{gathered}
\end{equation}

where $\mathbf{u(x},t)$ represents the velocity of the suspension, $n(\mathbf{x},t)$ is the cell concentration, $p_{e} (\mathbf{x},t)$ is the excess pressure, $\rho$ is the fluid density, and $\mathbf{g}$ is the gravitational acceleration. An additional term, $\Sigma (\mathbf{x},t)$, captures the impact of the swimming cells on the bulk fluid stress \cite{pedley1990} due to a combination of the swimming-induced stresslets, Batchelor stresses, and stress associated with rotational diffusion. In concentrated suspensions, the first term (the swimming-induced stresslets, $\Sigma^{(p)}$) contributes to the leading-order stress for the aggregate cell swimming \cite{pedley2010,pedley2010-2}, thus yielding:

\begin{equation}
\begin{gathered}\label{eqn:PAM:vortical-critical5}
   \nabla \cdot \Sigma = \mu \nabla^2 \mathbf{u} + \nabla \cdot \Sigma^{(p)}
  \end{gathered}
\end{equation}

where, $\mu$ is the fluid viscostiy.

\begin{figure}
	\begin{center}
		\includegraphics[height=6cm]{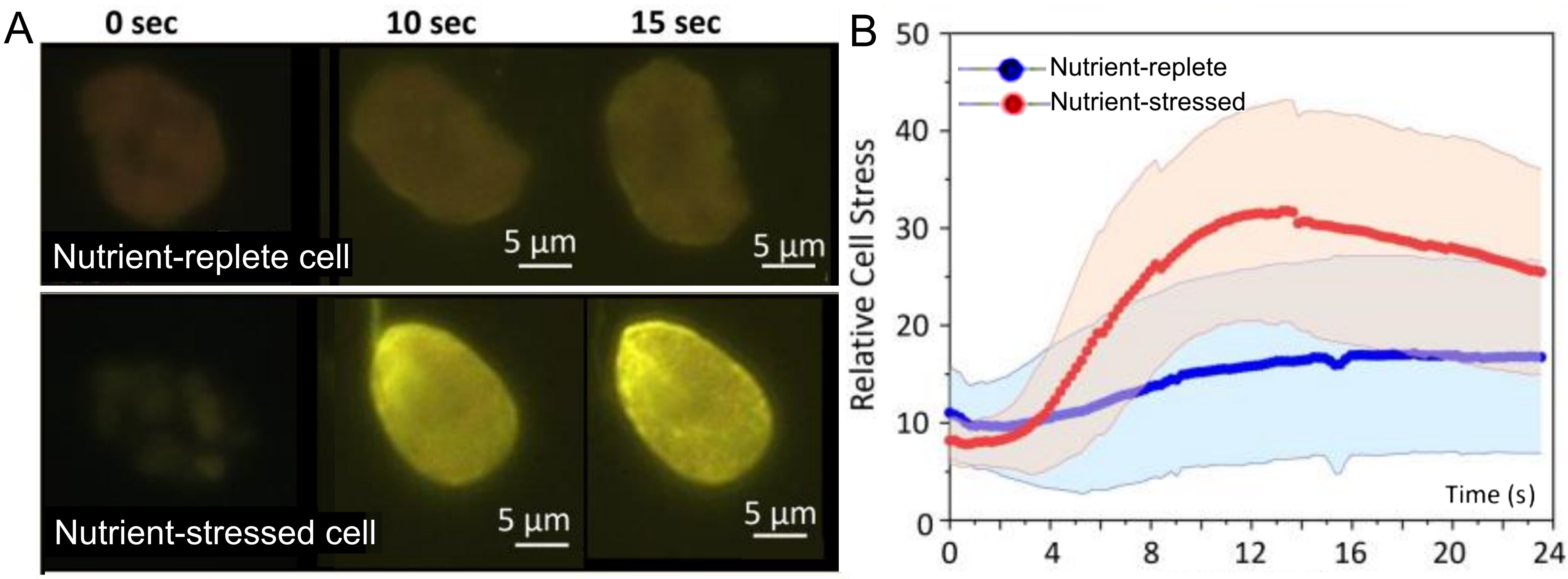}		
	        \caption{
	        {\bf Physiological stress drives active gravitactic alterations.} {\bf a} Epifluorescence intensity measures the production of reactive oxygen species (ROS) under control (nutrient-replete condition) and nutrient-depleted condition. ROS was visualized by staining cells with 5 $\mu$M CellROX Orange (Thermo Fisher). {\bf b} Time series of stress accumulation quantified as ROS production for cells after 180 h (blue) and after 396 h (red). Solid line represents arithmetic mean of analyzed cells and shaded regions represents mean ± s.d. Time represents the acquisition time under microscope. Adapted from \cite{sengupta2022}.
	   	}\label{fig:stress}
	\end{center}
\end{figure}

Since the timescale for bioconvection is much shorter than that of the cell growth, we can derive a conservation equation of the form: 

\begin{equation}
\begin{gathered}\label{eqn:PAM:vortical-critical6}
   \frac{\partial n}{\partial t} = - \nabla \cdot \left[n(\mathbf{u + v}) - D \cdot \nabla n \right]
  \end{gathered}
\end{equation}

where the flux terms represent advection by the flow $\mathbf{u}$, drift relative to the flow with mean swimming velocity $\mathbf{v(x,}t)$, and the swimming diffusion tensor $D (\mathbf{x},t)$. Descriptions for individuals are either phenomenological or derived from external torques coupled to low–Reynolds number fluid mechanics \cite{bees2020}. Appropriate forms of $\mathbf{v(x,}t)$ and $D (\mathbf{x},t)$ for each taxis are described in detail in the References \cite{hill2002,manela2003,hwang2014}. As initial conditions, a fully mixed suspension or a steady, horizontally uniform, vertical profile could be considered. 

Bioconvection can have important ramifications in both industrial and natural settings, and enable better understanding of the physico-chemical ecology and population-scale behavior of the microbial suspensions and emergent structures which they generate \cite{bees2014maths}. Considering the high concentration in cells required in applications like the the production of algal biodiesel, synthesis of metabolic byproducts within photobioreactors ($e.g.$, beta carotene), algae-derived nutrients, and algal H$_{2}$ production, precise understanding of bioconvection in these systems could help to optimize the energy budget, a key technological challenge \cite{kessler1985-2,bees2014maths,williams2014}. Beyond laboratory and industrial settings, bioconvection has been reported to be an important driver of molecular transport in natural ecosystems (Figure \ref{fig:bioconv} \textbf{e-f}), particularly in meromictic lakes which lack physical mechanisms for mixing and transport of heat and molecular cargo \cite{gulati2017book}. Combining field and laboratory studies, Sommer $et al.$ \cite{sommer2017} provided evidence that bioconvection mediates the eco-physiology of photosynthetic bacteria present in quiescent water bodies like the alpine Lake Cadagno (Tecino, Switzerland). The study revealed that bioconvection by \textit{Chromatium okenii}, a 10 $\mu$m-long sulphur-metabolizing motile purple bateria, was adequate to sustain mixed layers within the lake. The bacteria, found to be 15–27\% denser than water, swims at $\sim$2-3 body lengths per second to generate a mean theoretical dissipation rate of $1.0 \pm 1.5 \times 10^{-10}$ W/kg, which is $\sim$45\% of the bacterial energy input rate.

\section{Emerging concepts in planktonic active matter}
Planktonic active matter is inherently adaptive: depending on the environmental cues, among others nutrients, light, temperature and turbulence levels, species can respond and adapt to the changes in their environment \cite{thomas2012,ackermann2015} using exquisite biophysical mechanisms which are recently being discovered \cite{sengupta2017,carrara2021,sengupta2020,sengupta2022}. Today, oceans are undergoing a major makeover due to changes in a range of biotic and abiotic factors, thus understanding phytoplankton behaviour and response under these dynamic settings is a need of the hour \cite{behrenfeld2006,vandewaal2020}. In this section, we will briefly cover a selection of emerging concepts in plankton ecology aimed toward a mechanistic understanding of the \textit{smart} behavioural traits \cite{colabrese2017} which often emerge as a response or adaptation to changing environments.

\subsection{Active shape-shifting microplankton}

The ability of phytoplankton to rapidly adapt morphology allows cells to dynamically switch the direction of migration, and redistribute along the vertical ocean column as conditions turn inclement \cite{sengupta2017}. Rapid polymorphism, achieved via fine-scale control of the cellular morphology, is an important adaptive trait that phytoplankton harness as a potential bet-hedging strategy. Evolutionary bet-hedging involves a trade-off between the mean fitness and the temporal variance of fitness, such that phenotypes with reduced arithmetic mean fitness may be at a selective advantage under fluctuating environmental conditions \cite{seger1987,philippi1989}. The diversification of migration strategies, such as the one observed in raphidophytes experiencing turbulent cues \cite{sengupta2017} or due to prolonged nutrient limitation \cite{sengupta2022}, qualitatively suggest the possibility of bet-hedging strategy in action, however quantitative links are yet to be established. Cell shape and size are known to directly influence an organism’s ecophysiology and fitness \cite{ryabov2021}, thus the expansion of the phenotypic space over short timescales suggests a functional links between morphodynamics and emergent diversity, which are yet to be fully understood.

\begin{figure}
	\begin{center}
		\includegraphics[height=6cm]{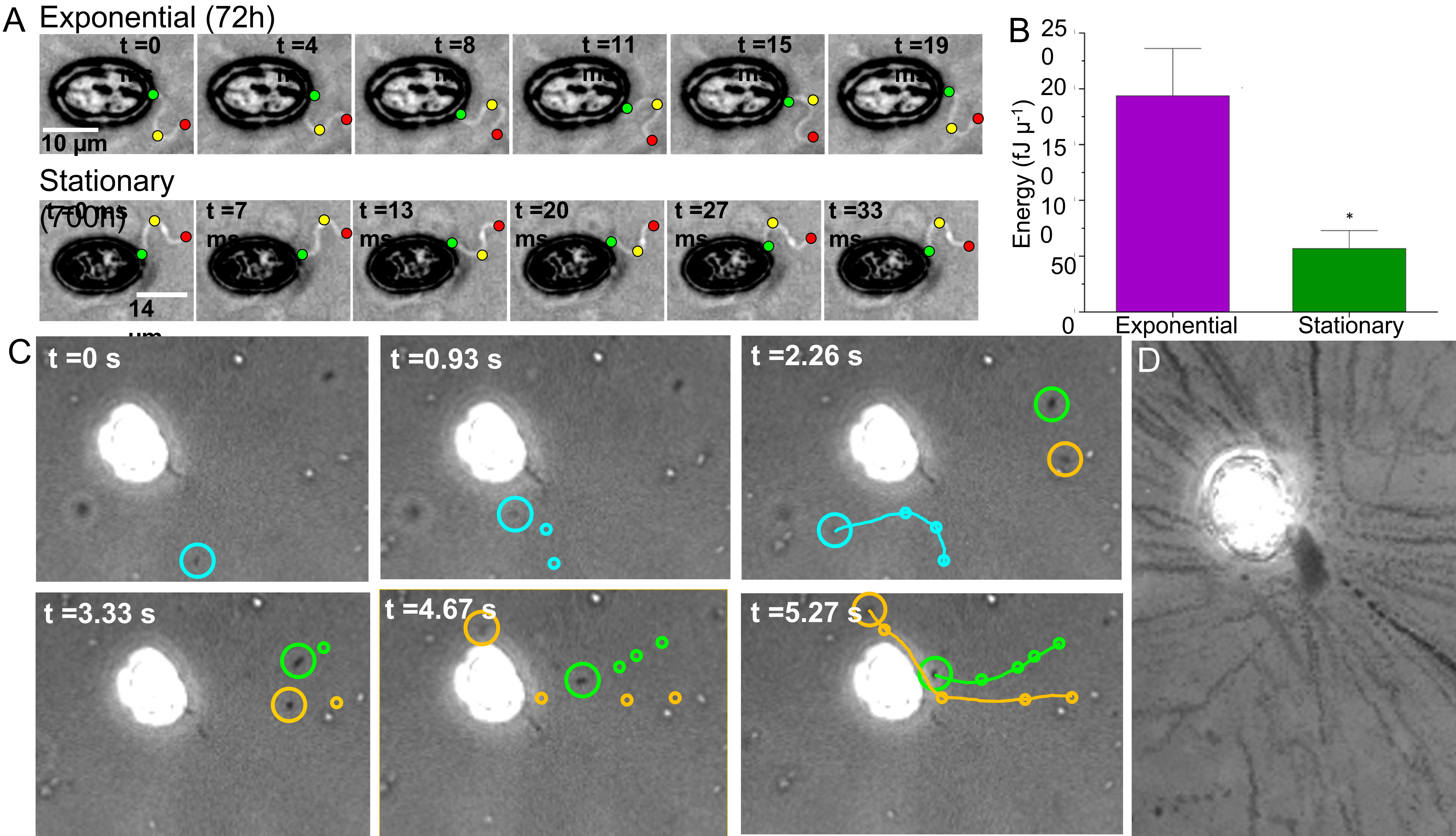}		
	        \caption{
	        {\bf Microplankton trophic shifts and adapation to nutrient-stresses.}
	  {\bf a} Flagellar motion and associated energy due to flagellar beating of cells during exponential growth phase and stationary phase. Videos were acquired at frame rates ranging between 200 to 350 frames/sec (sufficiently high to cover 
successive flagellar strokes). {\bf b} Flagellar energy for the exponential and stationary phases are plotted using
bar graphs showing the mean ± s.d, and asterisk indicates statistical difference. {\bf c} Trajectories of preys (beads and non-motile bacteria) reveal feeding currents generated by microplankton cells after prolonged nutrient limitation. {\bf d} A time-integrated micrograph shows trajectories of multiple bacterial cells trapped in the feeding current generated by the microplankton. Adapted from Ref. \cite{sengupta2022}.
	    }\label{fig:MD:acrossscales}
	\end{center}
\end{figure}

\subsection{Tuning gravitaxis via intracellular reorganization}

The ability to migrate vertically along the water column, either by swimming or by buoyancy regulation, allows phytoplankton critical access to light and nutrient-rich patches. Under low nutrient concentrations, movement of cells relative to their surrounding increases the flux of nutrient molecules (relative to pure diffusion), thus rendering competitive advantage to motile species under low nutrient settings \cite{ross2007,margalef1978mandala,schuech2014}. Leveraging migration along the vertical ocean column, phytoplankton may explore a viable acclimatization alternative to the non-migratory trait shifts, provided the benefits outweigh the costs associated with swimming. Consequently, whether a population continues migration to optimize nutrient uptake (relying on energy reserves), or depends on chanced encounters with ephemeral nutrient patches is a long-standing conundrum \cite{grover2009,grover2017}. Recent studies indicate that phytoplankton harness lipid droplets--energy-storing organelles--which form under nutrient-limitation--as biomechanical triggers to regulate swimming properties under nutrient limitation \cite{sengupta2022}. By actively controlling the position and size of the droplets, cells decide whether to swim up or down. Other intracellular organelles including chloroplasts, carbohydrate granules, or biomineralized globules may as well modulate gravitaxis. Future studies focusing on the role of intracellular, reconfigurable organelles could shed light on the biomechanics of adaptive gravitaxis under ecologically relevant settings. 

\subsection{Repurposing flagellar beating}

Flagellar dimensions and beating patterns can alter the effective drag force acting on swimming cells, in addition to altering the cell length and the aspect ratio \cite{elgeti2015,visser2006,guasto2012}. For instance, a swimming cell executing a breaststroke-like motion ($e.g.$, Chlamydomonas or Dunaliella), can be represented by different effective aspect ratios over a beat cycle, depending whether the flagella occupy positions which are protruding out toward the front of the cell (along the long axis of the cell body), or occupy complementary positions where the flagella protrude sideward resulting in a lower aspect ratio. To what extent phytoplankton actively alter flagellar beating patterns to generate desired orientational dynamics (by varying the effect drag force and the cell aspect ratio) is an open question. Recent studies indicate that phytoplankton may repurpose their flagellar beating under ecological constraints \cite{sengupta2022}. For instance, under nutrient-replete conditions, motile species utilize flagella for generation of propulsion force, whereas under nutrient-limited settings the same species can employ flagells for two specific tasks: (a) to anchor to available surfaces, and (ii) to generate feeding currents to support the shift from phototrophic to mixotrophic foraging mode (Figure \ref{fig:MD:acrossscales}). Thus, it remains to be seen if the diversifcation of the flagellar functions is genetically encoded in species, and if so, how do the biomechanical properties emerge in order to optimally support the desired cellular requirements under different environmental constraints. 

\begin{figure}
	\begin{center}
		\includegraphics[height=10cm]{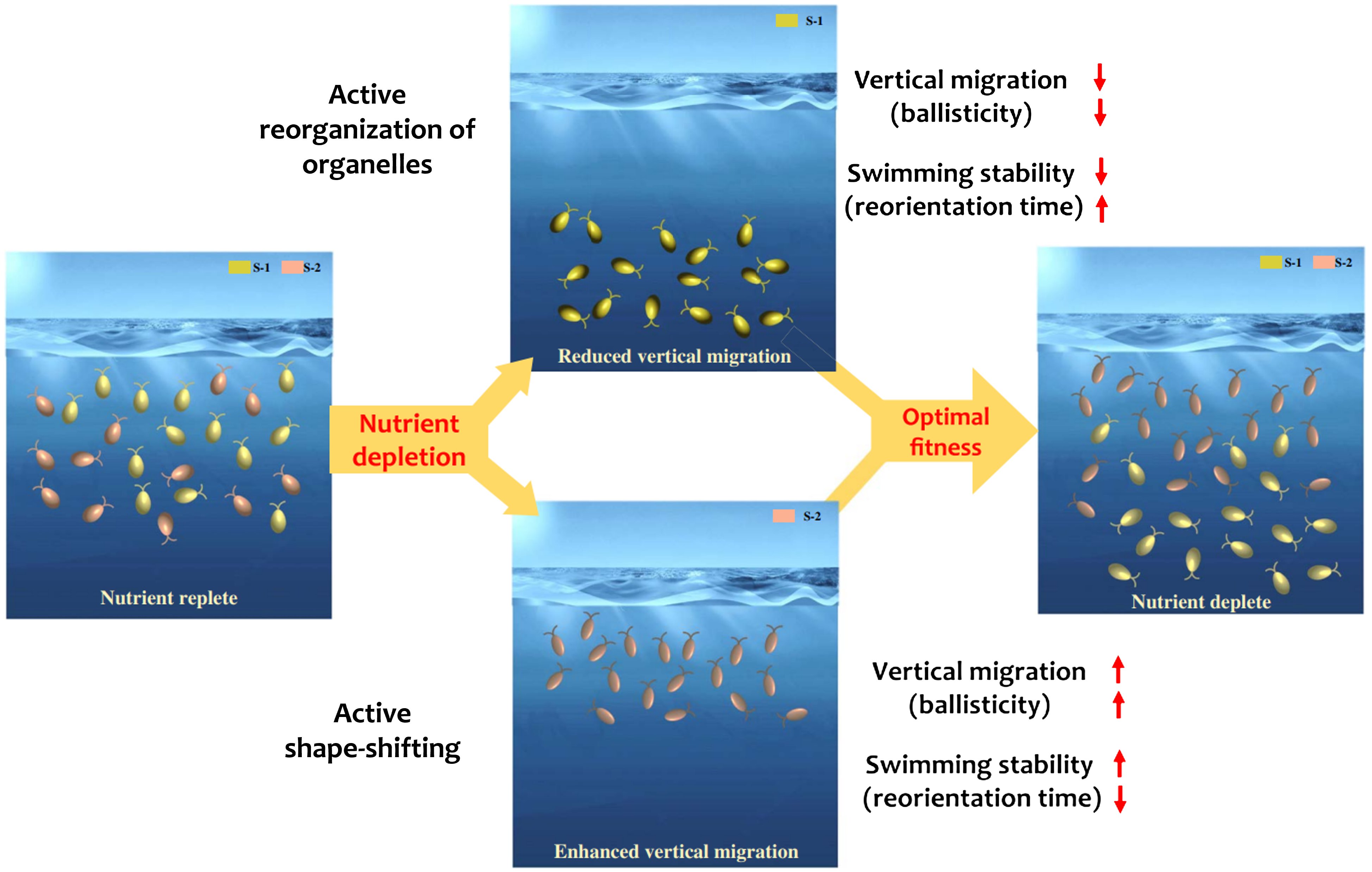}		
	        \caption{
	        {\bf Intelligent navigation toward niche expansion along the ocean column.}
	   	Microplankton leverage energy storing organelles (e.g., lipid droplets)  along with morphological changes to dynamically adapt active swimming strategies. This allows planktonic populations to enhance fitness under stressful environmental conditions, shown here for nutrient-limited settings. Different strains can adapt swimming behaviour differently, for instance, enhance motility (high ballisticity and low reorientation time) versus lowered motility. Shown here, strain S-1 harnesses lipid droplets to suppress vertical migration under nutrient limitation, whereas the sister strain (S-2) strengthens negative gravitaxis facilitated by the morphological change (despite lipid production). S-1 cells thus conserve energy by switching from ballistic to diffusive swimming in deeper waters; S-2 cells on the other hand, reduce energy requirements by lowering the viscous losses (due to modified shape). The strain-specific behavioral and physiological changes allow species to expand their ecological niche along the vertical water column, thus enhancing their chances of survival under stressful conditions. Adapted from \cite{sengupta2022}.
	    }\label{fig:nutrients}
	\end{center}
\end{figure}

\subsection{Role of reactive oxygen and nitrogen species}

Reactive oxygen and nitrogen species (RONS), although cytotoxic, serve as crucial stress signalling molecules at low concentrations within phytoplankton species, across different stages of growth and development \cite{kim2005,dautreaux2007,portune2010,diaz2018}. Cells produce RONS under diverse conditions involving oxidative stress \cite{lesser2006}. RNS and ROS can regulate plant and algal responses, both under normal physiological conditions and in response to biotic and abiotic stress, including variations in temperature, salinity, pH, light, and physical damages, through cross-talk between oxidative and nitrosative signalling \cite{sengupta2017} (Figure \ref{fig:stress}). RONS bursts are commonly observed in marine eukaryotes, including raphidophytes \cite{kim2008}. Recent evidence suggest that generation of physiological stress (resulting in RONS) is the mechanism responsible for the switch in the direction of migration in the marine raphidophytes \cite{sengupta2017,carrara2021}. Cells integrate information on the ROS within timescales as short as tens of seconds, allowing cells to rapidly alter swimming direction in response to turbulence. The results provide an exciting new perspective on the ecological interrelation between gravitaxis and physiological stress, and the potential role of RONS in sensing turbulence and gravity forces intensities in natural environments.

 \section{Conclusions} 

Planktonic active matter possess unique abilities to adapt motility strategies using a combination of morphological, flagellar, intracellular changes. These chnages allow the cells to enhance fitness and chances of survival under stressful conditions. Diversification of traits, both physiological and behavioral, enhance the chances of survival of species across a wide range of environmental settings. This is mediated via exquisiite biophysical and biomechanical changes, many of which remain to be understood, or even discovered. One of the main future directions of investigation will be to foster understanding of the active sensing and communication mechanisms in phytoplankton for flow and turbulence-sensing under constraints of nutrients and other resources including light (Figure \ref{fig:nutrients}). Understanding how phytoplankton adapt and survive the rapidly evolving nutrient landscapes of today’s oceans remains a crucial challenge. Accurate prediction of harmful blooms due to the raphidophytes, and more broadly, the cascading biogeochemical implications of phytoplankton adaptations, will rely on mechanistic understanding of co-emerging behavioural and physiological responses of phytoplankton communities, and the interrelations therein. The concepts and perspectives covered in this chapter, though somewhat limited in scope relative to the vast richness of plankton and their active dynamics in real world settings, should offer tools and ideas to apply and advance mechanistic frameworks to capture phytoplankton behaviour and interactions under diverse environmental conditions; and open up quantitative approaches to assess impacts of multiple stressors which are increasingly plaguing today's aquatic ecosystems.

 \section{Acknowledgement} 
 AS thanks the Luxembourg National Research Fund's ATTRACT Investigator Grant (Grant no. A17/MS/ 11572821/MBRACE) and CORE Grant (C19/MS/13719464/TOPOFLUME/Sengupta) for supporting this work. This work wouldn't have been possible without numerous enriching discussions with collaborators and colleagues, some of whom I have been lucky to work with as a co-author over the last many years. Finally, I would like to thank the Editors of the book for their kind invitation to contribute this chapter on Planktonic Active Matter.

\clearpage

\bibliography{main}

\end{document}